\documentclass[12pt,letter]{article}

\usepackage{scicite}
\usepackage{times}
\usepackage{graphics,graphicx}

\usepackage{amsmath, amssymb}
\usepackage{url}
\usepackage{multibib}
\usepackage{hyperref}

\usepackage{color}

\usepackage{multirow}   

\def\aap{Astron. Astrophys.}                
\newcommand{\aj}{Astron. J.}

\newcommand{\apjs}{Astrophys. J. Suppl. Ser.}
\newcommand{\apj}{Astrophys. J.}
\newcommand{\apjl}{Astrophys. J.}
\newcommand{\mnras}{Mon. Not. R. Astron. Soc.}
\newcommand{\nat}{Nature}

\newcommand{\pasp}{Publ. Astron. Soc. Pac.} 

\newcommand{\araa}{Annu. Rev. Astron. Astrophys.}

\newcommand{\ssr}{Space Sci. Rev.}
\newcommand{\prd}{Phys. Rev. D}

\newcommand{\degr}{\hbox{$^{\circ}$}}

\newcommand\fdg{\mbox{$.\!\!^\circ$}}%
\newcommand\arcsec{\mbox{$^{\prime\prime}$}}%

\newcommand{\maxi}{MAXI~J1820+070}

\newcommand*{\doi}[1]{\href{http://dx.doi.org/#1}{doi:#1}}

\newcommand{\unitvec}[1]{\mbox{\boldmath $\hat{#1}$}}

\newenvironment{sciabstract}{%
\begin{quote} \bf}
{\end{quote}}

\title{Black hole spin--orbit misalignment in the x-ray binary MAXI~J1820+070}

\author
{Juri Poutanen$^{1,2,3\ast}$, Alexandra Veledina$^{1,3,2}$, Andrei~V.~Berdyugin$^{1}$, \\
Svetlana~V.~Berdyugina$^{4}$, Helen Jermak$^{5}$, Peter G.~Jonker$^{6,7}$, Jari J.~E.~Kajava$^{1,8}$, \\
Ilia A. Kosenkov$^{1}$,   Vadim Kravtsov$^{1}$, Vilppu Piirola$^{1}$, Manisha Shrestha$^{5,9}$, \\
Manuel A.~P.~Torres$^{10,11}$,  Sergey S.~Tsygankov$^{1,2}$}

\date{}

\begin{document}

\baselineskip 24pt
\baselineskip 12pt

\maketitle 

\noindent
1: Department of Physics and Astronomy, FI-20014 University of Turku, Finland\\
2: Space Research Institute  (IKI) of the Russian Academy of Sciences, 117997 Moscow,  Russia\\
3: Nordic Institute for Theoretical Physics (Nordita), KTH Royal Institute of Technology and Stockholm University, SE-10691 Stockholm, Sweden\\
4: Leibniz-Institut f\"{u}r Sonnenphysik, 79104 Freiburg, Germany\\
5: Astrophysics Research Institute, Liverpool John Moores University,  L3 5RF Liverpool, United Kingdom\\
6:  Department of Astrophysics, Institute for Mathematics, Astrophysics and Particle Physics (IMAPP), Radboud University, NL-6500 GL Nijmegen, The Netherlands\\
7: Space Research Organisation of the Netherlands (SRON), Netherlands Institute for Space Research,  NL-2333, CA Leiden, The Netherlands\\
8: Centro de Astrobiolog\'ia, Villanueva de la Ca\~nada, S-28692 Madrid, Spain\\
9: Department of Physics \& Astronomy, University of Denver, Denver, CO 80208, USA\\
10: Instituto de Astrof\'isica de Canarias, E-38205 La Laguna, Tenerife, Spain\\
11: Departamento de Astrof\'isica, Universidad de La Laguna, E-38206 La Laguna, Tenerife, Spain\\

\normalsize{$^\ast$Corresponding author. E-mail:  juri.poutanen@utu.fi.}


\begin{sciabstract}
The observational signatures of black holes in x-ray binary systems depend on their masses, spins, accretion rate and the misalignment angle between the black hole spin and the orbital angular momentum. We present optical polarimetric observations of the black hole x-ray binary MAXI~J1820+070, from which we constrain the position angle of the binary orbital axis. Combining this with previous determinations of the relativistic jet orientation axis, which traces the black hole spin, and the inclination of the orbit, we determine a lower limit of $40\degr$ on the spin-orbit misalignment angle. The misalignment has to originate from either the binary or black hole formation stage. If other x-ray binaries have similarly large misalignments, these would bias measurements of black hole masses and spins from x-ray observations. 
\end{sciabstract}


Black holes can be characterized with just two parameters: mass and spin. 
When a black hole resides in a binary system, accreting material from a companion donor star through the accretion disk, there are additional parameters that determine its observational signatures: the mass accretion rate  and the misalignment angle between the black hole spin and the orbital axis. 
Standard methods to measure black hole spin from x-ray observations -- iron line spectroscopy\cite{Reynolds14} or modeling of the accretion disk spectrum\cite{McClintock14} -- assume that the misalignment angle is small. 
Conversely, the standard interpretation of low-frequency quasi-periodic oscillations in x-ray and optical observations of black hole x-ray binaries as precession of the  accretion disk\cite{FB07,IDF09,VPI13}, requires the assumption that the misalignment angle is non-negligible. 
Substantial misalignment is theoretically predicted for x-ray binaries that received high velocities during formation\cite{Atri2019}.
The misalignment angle must be inherited from the formation process, because it can only decrease when the black hole is accreting\cite{maccarone2002}.  
Gravitational wave observations of merging black holes have detected signatures of orbital precession\cite{ligo2021second_catalogue} indicating non-zero misalignment in these systems\cite{apostolatos1994}, though they might not be representative of the wider population. 

Measuring the misalignment angle in x-ray binaries requires determining the three-dimensional orientation of the black hole spin and orbital axis.  
Accreting black holes often show relativistic jets, which are launched along an axis determined by the black hole spin direction\cite{McKinney13}.
The jet inclination angle can be directly obtained in some cases  from radio observations\cite{Mirabel99}, whereas the jet position angle can be measured using either radio or x-ray imaging. 
Combining these two angles allows the black hole spin orientation to be determined. 
Orbital  parameters, such as period and orbital inclination, can be determined using spectroscopic measurements of radial velocities of the donor star taken during quiescence, the stage at which accretion to the black hole is reduced and optical emission is not dominated by the accretion disk,  through orbital modulation of the optical photometry, and using constraints from the presence or absence of x-ray and optical disk eclipses\cite{Torres20}. 
 
The black hole x-ray binary \maxi\ was discovered as a transient x-ray source on 2018 March 11\cite{kawamuro18}.
X-ray quasi-periodic oscillations, detected shortly after the discovery, were observed for more than 100 days\cite{Stiele20}.
Ejections of material traveling at relativistic velocities have been observed from this source in both radio and x-rays, indicating the jet inclination (measured from the line-of-sight) is $i_{\rm jet}=63\degr\pm 3\degr$ and the position angle (measured on the plane of the sky from North to East) is $\theta_{\rm jet}=25\fdg1\pm 1\fdg4$\cite{Atri20,Bright20,Espinasse20}.
Both angles were determined to be stable over the observed duration of the outburst. 
The orbital inclination has been constrained to the range $66\degr<i_{\rm orb}<81\degr$  by the lack of x-ray eclipses and the detection of grazing optical eclipses\cite{Torres20}. 
To determine the orientation of the orbital axis requires one further parameter:  the orbital position angle $\theta_{\rm orb}$.

We monitored \maxi\ in the optical $B$, $V$, and $R$ photometric bands using the Double Image Polarimeters\cite{PBB14,Piirola21} during the 2018 outburst and in quiescence. 
We obtained the source intrinsic linear polarization by subtracting the foreground interstellar polarization, measured from nearby field stars.  
During the outburst, when the relativistic jets were detected at radio frequencies, the intrinsic linear polarization degree (PD) in the $V$ and $R$ bands reached 0.5\% at a polarization angle (PA, also measured from North to East) of 23\degr--24\degr, which coincides with the jet position angle within the uncertainties\cite{Veledina19,Kosenkov20}.
After the source faded in the x-rays, the PD increased by a factor of 5--10 and the PA changed by 40\degr$\pm$4\degr\ to $-$17\degr$\pm$4\degr\ (Table~\textbf{S1} and Fig.~\textbf{1})\cite{SuppMaterial}.
This increase in PD is most prominent in the $B$-band, which also has the highest PD in the range 1.5--5\%; the $R$-band polarization changes from 0.4\% to 2\%.
The PA is most precisely determined in the $B$-band,  which also shows the least variability, with the mean being  $\langle \mbox{PA}\rangle=-19\fdg 7\pm 1\fdg2$.

\begin{center} 
 \includegraphics[width=0.6\textwidth]{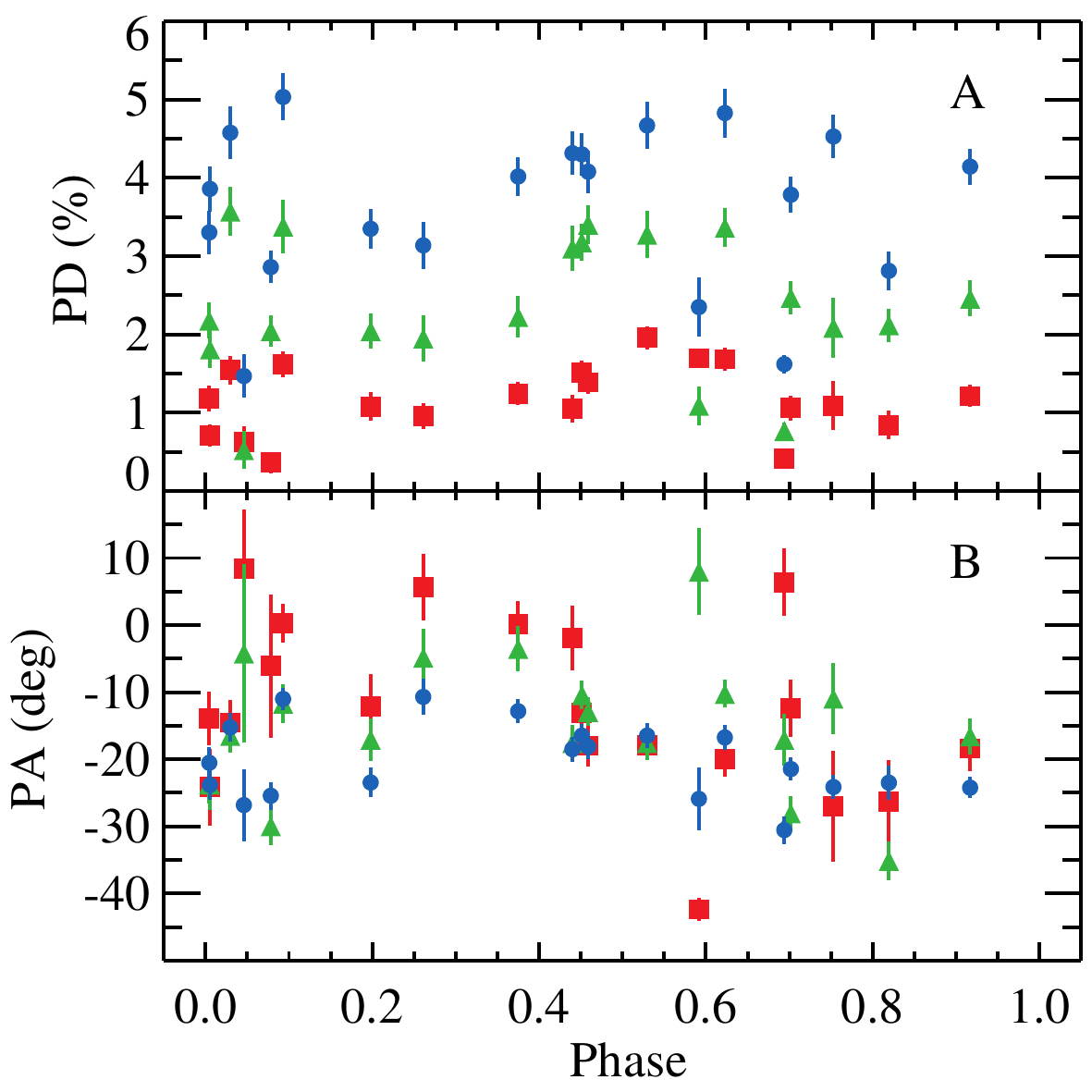}
 \end{center}

\noindent {\bf Fig. 1.} \textbf{Observed optical polarization properties of \maxi.}
{\small \textbf{(A)} Intrinsic polarization degree (PD) and \textbf{(B)} polarization angle (PA) of \maxi\ in quiescence are shown as a function of orbital phase [using a published ephemeris\cite{Torres19}]. 
The intrinsic values were obtained from the observed ones by subtracting the foreground interstellar polarization, measured from nearby field stars.
Blue circles, green  triangles and red squares correspond to $B$, $V$, and $R$ bands, respectively, with the error bars showing the 68\% confidence level.
Polarization is strongest in the $B$ band and weakest in the $R$ band, while the angle does not change substantially.}
\bigskip  

We identify three properties of the quiescent-state polarization: it is strongest in the blue part of the optical, with approximate dependence on frequency $\nu$ as PD$({\nu})\propto\nu^{3}$ (Table~\textbf{S1}, Fig.~\textbf{2}), the PD remains high in the range 0.5--5\% and the PA is stable. 
The PA undergoes apparently stochastic variations with an amplitude of $<$10\degr\ with no dependence on the orbital phase. 
These  properties constrain the mechanism of the polarized emission. 
We modeled  broadband photometric data obtained with the Liverpool Telescope and the \textit{Swift} Ultraviolet and Optical Telescope (UVOT) together with the polarized fluxes (Fig.~\textbf{2}).
We decompose the total spectral energy distribution into three components: a companion star (contributing $\sim25\%$ to the $R$-band flux), an accretion disk with inner temperature $T_{\rm d}\approx6,200$~K and inner radius $R_{\rm d}\approx6\times10^{10}$~cm, and an additional ultraviolet (UV) component with blackbody temperature $T_{\rm bb}\approx15,000$~K and radius $R_{\rm bb}\approx 9\times10^{9}$~cm (Table~\textbf{S4}). 
The properties of the polarized flux are consistent with being produced by the UV component with constant PD of 5--8\%. 

\begin{center} 
\includegraphics[width=0.6\columnwidth]{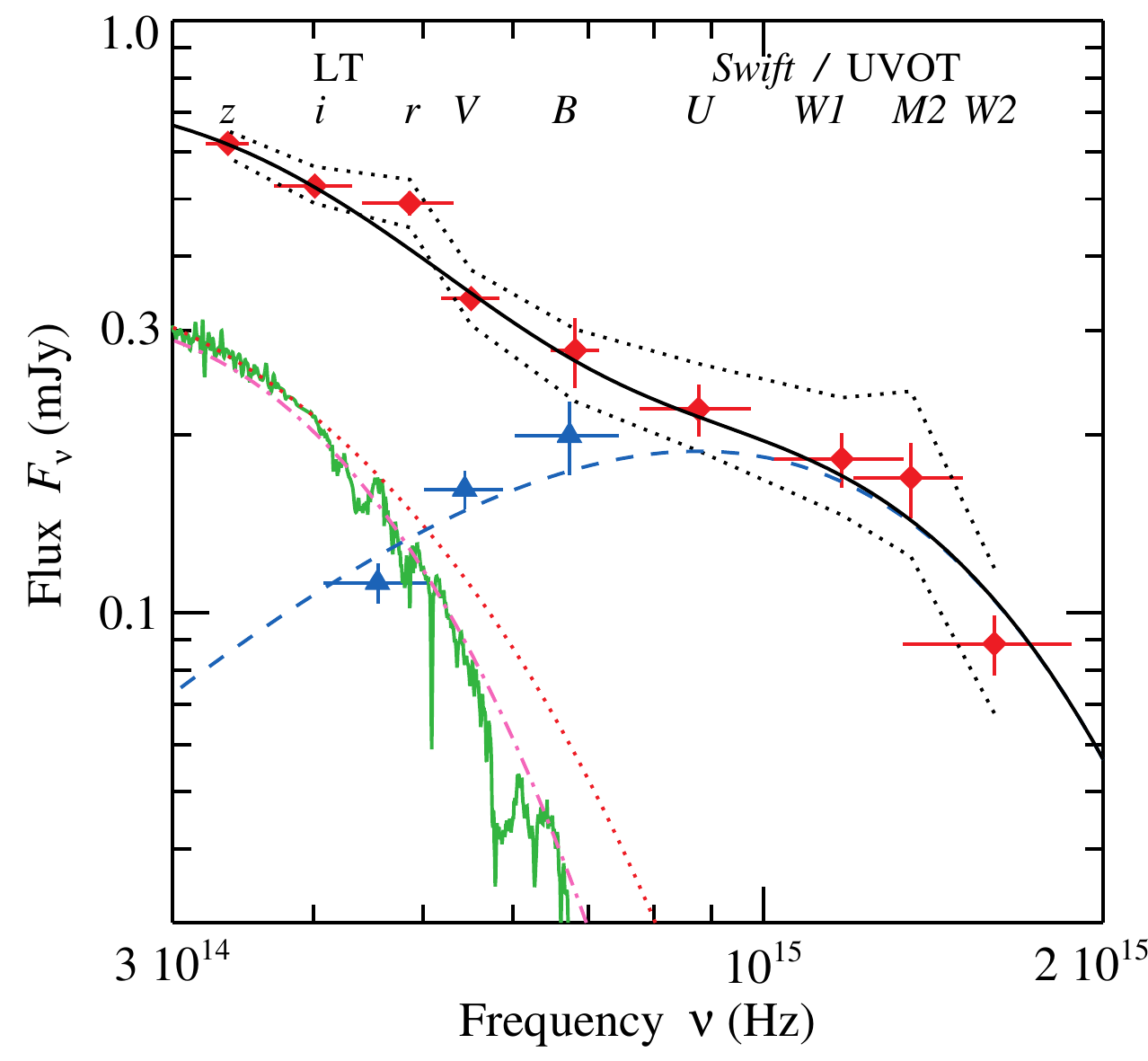}
 \end{center}

\noindent {\bf Fig. 2.} \textbf{Spectral energy distribution (SED).} 
{\small The average SED of \maxi\ (red diamonds) as observed by the Liverpool Telescope and \textit{Swift}/UVOT telescope in July 2020 and corrected for reddening with color excess $E(B-V)=0.29$. 
The photometric bands are indicated at the top of the figure.
The black dotted lines give the lower/upper limits on the flux for lower/higher extinction with $E(B-V)=0.25$ and 0.325, respectively. 
The polarized flux divided by the best-fitting model polarization degree $P_{\rm UV}=0.055$ (i.e. multiplied by a factor of $\sim$18) is shown with blue triangles.  
Error bars show 68\% confidence level.
The black line gives the total model flux consisting of the companion star modeled as a blackbody (pink dot-dashed line), accretion disk (red dotted) and a hot blackbody (blue dashed).  
The spectrum of a K7 star\cite{Pickles98} is shown as the solid green line for comparison. }
\bigskip

The jet cannot be the source of the polarized emission because its optically thin synchrotron spectrum is red, inconsistent with the observed blue  spectrum of polarized light.
Moreover, the PA is offset by about $40\degr$ from the jet position angle. 
The absence of detectable orbital variations in the PA excludes a hot spot origin.
An optically thick accretion disk is excluded by the high PD and blue spectrum. 
A potential source of the polarized emission is scattering of the accretion disk radiation in the hot optically thin and geometrically thick accretion flow close to the disk inner radius\cite{SuppMaterial,Narayan96}, which can also be responsible for the observed UV excess.
This mechanism would produce polarization parallel to the meridional plane, i.e. the plane formed by the orbital axis and the direction towards the observer.
Another possibility is dust scattering, thought to be responsible for blue polarized spectra observed from accretion disks around some supermassive black holes\cite{Webb93}.
The presence of dust in quiescent-state black hole x-ray binaries has been inferred from the detection of the mid-IR excess in two systems\cite{Muno06}.
If dust is located within a flattened envelope, in the wind around the accretion disk,  or in a circumbinary disk, the resulting polarization vector would also be parallel to the meridional plane. 
However, if dust forms an extended, approximately spherical structure at high elevation above the accretion disk, the polarization would be  perpendicular to the meridional plane. 
We consider the latter scenario to be implausible, as a nearly spherical envelope cannot produce the high observed PD.
A dust scattering mechanism would not explain the UV excess, because the disk does not emit in that range and hence there are no photons to be scattered by the dust. 

\begin{center} 
\includegraphics[width=0.7\columnwidth]{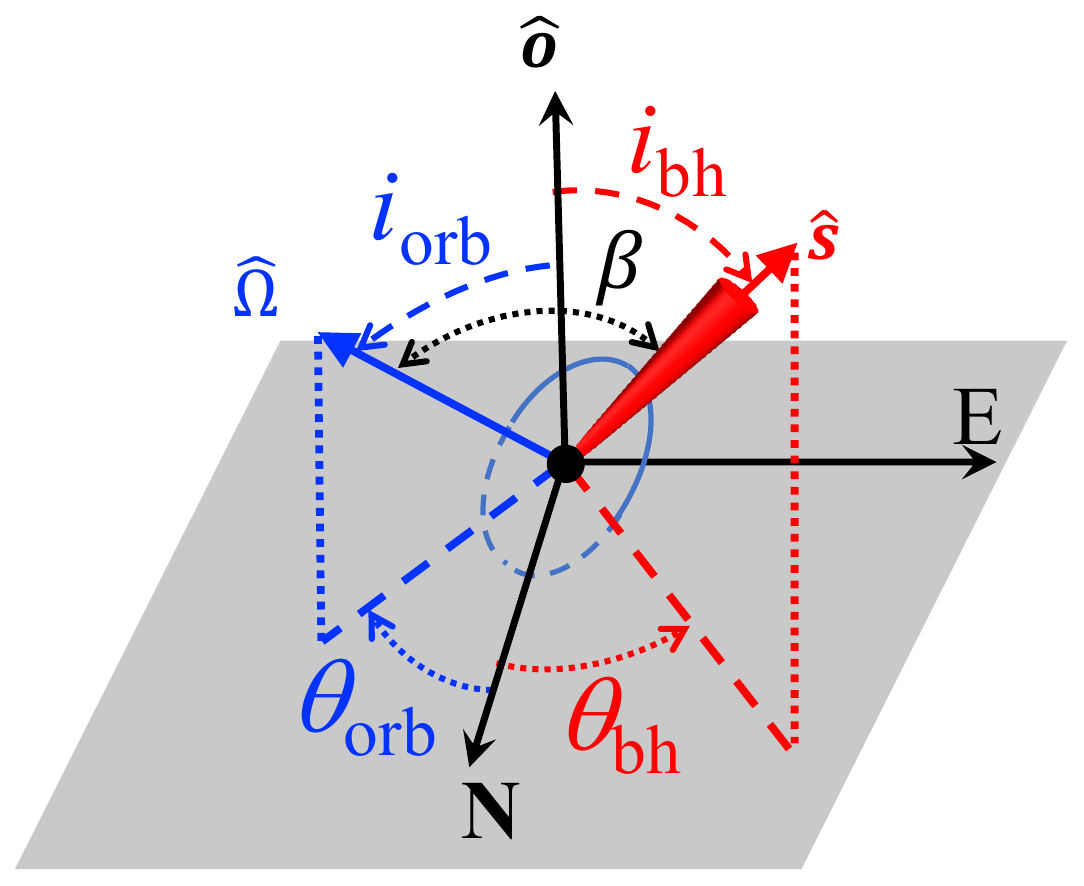}
 \end{center}
 
\noindent {\bf Fig. 3.} \textbf{Geometry of the system from the observer's perspective.} 
{\small The grey plane is the plane of the sky, labelled with North and East axes, perpendicular to the line of sight towards the observer $\unitvec{o}$. 
The angles between the line of sight and the vectors of the orbital angular momentum $\unitvec{\Omega}$ and the black hole spin $\unitvec{s}$ are the inclinations  $i_{\rm orb}$  and $i_{\rm bh}$. 
The corresponding position angles $\theta_{\rm orb}$ and $\theta_{\rm bh}$ are the azimuthal angles projected onto the sky, measured from North to East. 
The misalignment angle $\beta$ is defined as the angle between $\unitvec{s}$ and $\unitvec{\Omega}$. 
The red cone indicates the jet and the blue ellipse indicates the companion star orbit around the black hole, which is at the coordinate center. }
\newpage

Independent of the spectral modeling and geometry of the emission, the stability of the PA (most evident in the $B$-band, Fig.~\textbf{1})  over the orbital phase suggests that the polarization is related to the orbital axis, either parallel or perpendicular to it.
Hence, the observed PA provides information about the position angle of the orbital axis.
The misalignment angle $\beta$ can be determined from 
\begin{equation} \tag{1}
    \cos\beta = \cos i_{\rm bh} \cos i_{\rm orb} + \sin i_{\rm bh} \sin i_{\rm orb} \cos \Delta,
\end{equation} 
where $i_{\rm bh}$ is the inclination of the black hole spin vector (measured from the line-of-sight) and $\Delta=\theta_{\rm bh}-\theta_{\rm orb}$ is the difference between the position angles of the black hole spin vector $\theta_{\rm bh}$ and the orbital angular momentum $\theta_{\rm orb}$ (the geometry is illustrated in Fig.~\textbf{3}). 
If the black hole spin vector is directed along the southern approaching jet, then its inclination $i_{\rm bh} = i_{\rm jet}=63\degr\pm 3\degr$ and its position angle is $\theta_{\rm bh}=180\degr+ \theta_{\rm jet} =205\fdg1\pm 1\fdg4$\cite{Atri20,Bright20,Espinasse20}. 
The smallest misalignment $\beta\approx 42\degr$ is achieved when the orbital spin is also directed South at $\theta_{\rm orb}=\langle \mbox{PA}\rangle$+180\degr=$160\fdg3\pm1\fdg2$ (because the PA has an ambiguity of 180\degr), at the inclination $i_{\rm orb}\approx73\degr$. 
The probability distribution for $\beta$ in this case is shown in Fig.~\textbf{4}.  
The radial velocity measurements\cite{Torres20} do not differentiate between orbital inclinations $i_{\rm orb}$ and $180\degr-i_{\rm orb}$, so there is a second solution with $i_{\rm orb}\approx107\degr$ and $\beta\approx 63\degr$.
If either the orbital angular momentum or the black hole spin is instead directed to the North, the black hole rotation is then retrograde, resulting in $\beta\approx 117\degr$ or 138\degr\ for the same two solutions for the orbital inclination as above.

If the polarization vector is perpendicular to the meridional plane, the orbital position angle can take values $\theta_{\rm orb}=\langle \mbox{PA}\rangle+90\degr$ or $\langle \mbox{PA}\rangle+270\degr$. 
This geometrical arrangement leads to nearly identical values for $\beta$ because the difference between jet position angle and $\langle \mbox{PA}\rangle$ is about 45\degr.
All possible cases for the orientations of the black hole and orbital spins, the resulting values for $\beta$  and the azimuthal angle of the black hole spin in the orbital plane are listed in Table~\textbf{S5}. 
Corresponding probability distributions are shown in Figs.~\textbf{S4} and \textbf{S5}.

The difference of $\approx45\degr$ between the jet position angle and the PA indicates $\gtrsim40\degr$ misalignment between the black hole spin and the orbital angular momentum.
This result is independent of modeling or geometric ambiguities, because it relies only on the observed difference between the polarization angle and  jet position angle.

During outbursts, when the matter reaches the black hole, this misalignment affects the innermost regions of the accretion disk.
For a non-zero spin, particles moving around the black hole in orbits tilted with respect to the black hole equatorial plane undergo precession at a rate that decreases with radius\cite{FB07}.
Hence, a tilted disk is subject to twist and warp. 
A high misalignment adds complications to the models of quasi-periodic oscillations observed in black hole x-ray binaries, which rely on precession of the inner parts of the accretion flow, implying the whole flow is misaligned by $2\beta$ from the orbital axis in some phases\cite{FB07}.
For $\beta\sim40\degr$ the inner parts of the accretion disk would need to become almost perpendicular to its outer parts.
Most models assume smaller misalignment angles, typically, $\beta\sim10\degr-20\degr$\cite{FB07,IDF09,VPI13}, although highly inclined possibilities with $\beta \sim 45\degr-65\degr$ have sometimes been considered\cite{Liska21}.

\begin{center} 
\includegraphics[width=0.6\columnwidth]{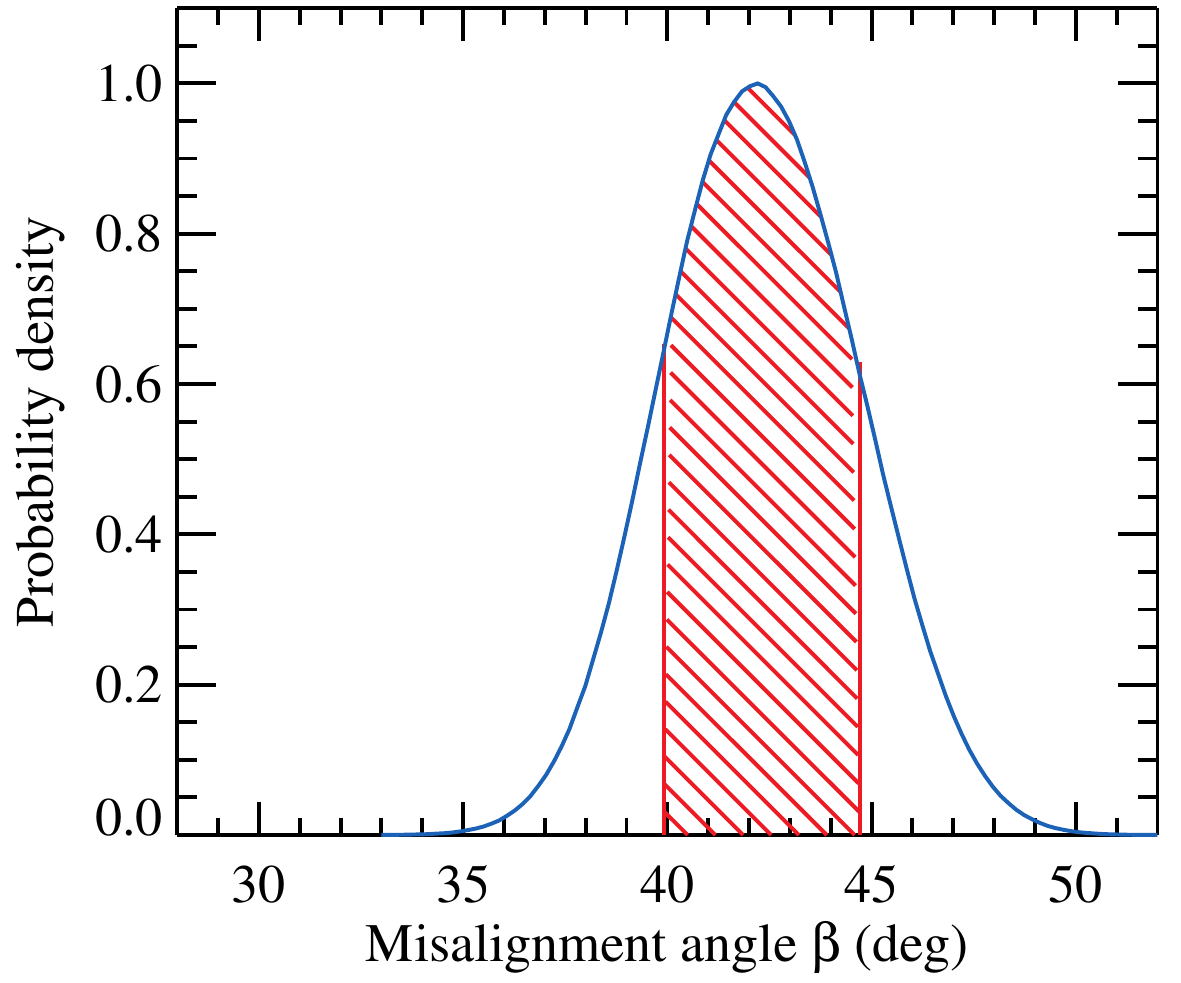}
 \end{center}
 
\noindent {\bf Fig. 4.} \textbf{Probability distribution function for the misalignment angle.} 
{\small The distribution normalized to the peak value is shown for the smallest misalignment angle possible. 
This case corresponds to the black hole spin directed along the southern approaching jet and the orbital spin being directed South at a position angle $\theta_{\rm orb}=\langle \mbox{PA}\rangle+180\degr$  and inclination $i_{\rm orb}\approx73\degr$.
The red hatched region corresponds to the 68\% confidence interval (i.e. between 16th and 84th percentiles of the posterior probability distribution). 
Distributions of $\beta$ for the other seven possible combinations of $\theta_{\rm orb}$, $i_{\rm orb}$ and $i_{\rm bh}$ are shown in Fig.~\textbf{S4}. }
\bigskip

High misalignment has previously been suggested based on observations of the  gamma-ray light curves produced by the jet in Cyg X-3\cite{Zdziarski18},  and differences between orbital and jet inclination angles are $15\degr$ in GRO~J1655--40\cite{maccarone2002} and $50\degr$ in V4641~Sgr\cite{Orosz01}  though the latter is highly uncertain. 
Misalignment has also been theorized, based on the inferred high kick velocities of x-ray binaries acquired during formation\cite{Atri2019}. 
For the black hole x-ray binary MAXI~J1820+070, the high misalignment was identified only after obtaining the constraints on the position angle of the orbital angular momentum $\theta_{\rm orb}$.
Without information on the binary plane orientation, we would have obtained only a lower limit on the misalignment angle in \maxi\ of $\gtrsim$5\degr, because the orbital inclination is only marginally different from the jet inclination.

Our results demonstrate the need to treat the misalignment angle as a free parameter when measuring black hole masses and spins. 
Assuming that the black hole spin and the orbital angular momentum are aligned introduces a systematic bias on  measurements\cite{Atri20,Torres20,Zhao21}.
A large misalignment angle is expected to drive precession of the binary orbital plane, altering the gravitational waves emitted during a subsequent merger event\cite{apostolatos1994}.
Evidence for orbital precession has been found from population properties of black hole mergers observed using gravitational waves\cite{ligo2021second_catalogue}.




\section*{Acknowledgements}
We thank K. Belczynski and A. Zdziarski for comments and suggestions.
A.V. thanks the International Space Science Institute (ISSI) in Bern, Switzerland, for providing the environment for stimulating discussions. 
This paper is based on observations made with the Nordic Optical Telescope, owned in collaboration by the University of Turku and Aarhus University, and operated jointly by Aarhus University, the University of Turku, and the University of Oslo, representing Denmark, Finland, and Norway, the University of Iceland and Stockholm University at the Observatorio del Roque de los Muchachos, La Palma, Spain, of the Instituto de Astrofisica de Canarias. 
We thank the NOT staff for their excellent support. 
DIPol-UF is a joint effort between University of Turku (Finland) and Leibniz Institute for Solar Physics (Germany). 
The Liverpool Telescope is operated on the island of La Palma by Liverpool John Moores University in the Spanish Observatorio del Roque de los Muchachos of the Instituto de Astrofisica de Canarias with financial support from the UK Science and Technology Facilities Council. 
{\bf Funding:} 
A.V. acknowledges support from the Academy of Finland grant 309308. 
J.P., A.V., V.K., and S.S.T. received funding from the Russian Science Foundation grant 20-12-00364.  
I.A.K. and V.K. thank Magnus Ehrnrooth foundation for support. 
S.V.B. acknowledges support from the ERC Advanced Grant HotMol ERC-2011- AdG-291659. 
M.A.P.T. acknowledges support from the State Research Agency (AEI) of the Spanish Ministry of Science, Innovation and Universities (MCIU) and the European Regional Development Fund (FEDER) under grant AYA2017-83216-P and via a Ram\'on y Cajal Fellowship (RYC-2015-17854). 
 {\bf Author contributions:} 
J.P. and A.V. initiated the project, performed modeling and led writing of the text.
A.V.B., V.P., and I.A.K. planned and performed polarimetric observations.
H.J., M.S. and M.A.P.T.  planned and executed observations at the Liverpool Telescope and V.K.  reduced the photometric data.
J.J.E.K. obtained the Swift observations and analysed the data together with S.S.T. 
P.G.J. and S.V.B. contributed to the interpretation of the results. 
All authors provided input and comments on the manuscript. 
 {\bf Competing interests:} We declare no competing interests. 
{\bf Data and materials availability:} 
The raw DIPol-UF data are available at Zenodo \url{https://zenodo.org/record/5767398}\cite{Berdyugin22}.
The Liverpool Telescope data are available from the data archive at \url{https://telescope.livjm.ac.uk/DataProd/} (proposal ID JQ20A01). 
The Swift observatory data used in this study are available through the NASA data archive at \url{https://heasarc.gsfc.nasa.gov} (ObsIDs 000106272**, with ** being 19--22,  24--26).  
Our software for computation of probability distribution of the misalignment angle and for modelling polarization properties of the hot accretion flow is available at  Zenodo \url{https://zenodo.org/record/5837857}\cite{Poutanen22}.



\clearpage

\renewcommand{\thepage}{S\arabic{page}} 
\setcounter{page}{1}

\includegraphics[width=0.25\textwidth]{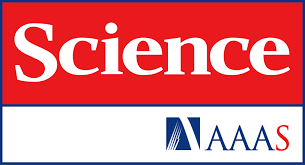}

\vspace*{1cm}

 \begin{center}
{\Large Supplementary Materials for} \\
\vspace*{1cm}

\textbf{\large Black hole spin--orbit misalignment in the x-ray binary MAXI~J1820+070}


\vspace*{1cm}

{Juri Poutanen$^{\ast}$, Alexandra Veledina, Andrei~V.~Berdyugin, Svetlana~V.~Berdyugina, \\
Helen Jermak, Peter G.~Jonker, Jari J.~E.~Kajava, Ilia A. Kosenkov,   Vadim Kravtsov, \\ 
Vilppu Piirola, Manisha Shrestha, Manuel A.~P.~Torres,  Sergey S.~Tsygankov}

\baselineskip 24pt
\baselineskip 12pt

\vspace*{0.5cm}
{$^\ast$ Corresponding author. Email: juri.poutanen@utu.fi}

\end{center}

\thispagestyle{empty}
 
 
\vspace*{2cm}

\noindent {\bf This PDF file includes:}\\

\noindent 
\hspace{1cm} Materials and Methods\\
\noindent\hspace*{1cm} Figs. S1 to S5\\
\noindent\hspace*{1cm} Tables S1 to S5\\
\noindent\hspace*{1cm} References 

\clearpage

 \section*{Materials and Methods}

\subsection*{DIPol-UF/NOT observations}

We have performed polarimetric observations of the black hole x-ray binary \maxi\cite{kawamuro18,denisenko18,Tucker18} both during the bright stages of the outburst in 2018 as well as in the quiescence. 
The data during the outburst\cite{Veledina19,Kosenkov20} were obtained with the Double Image Polarimeter-2 (DIPol-2) polarimeter\cite{PBB14} mounted on the Tohoku 60-cm telescope (T60) at Haleakala observatory, Hawaii. 
During the quiescence \maxi\ was observed with the DIPol Ultra Fast (DIPol-UF)\cite{Piirola21} at the 2.56-m Nordic Optical Telescope (NOT) in July 2019, April and July 2020, and July 2021 (Table~\ref{tab:dipol})\cite{Berdyugin22}.
Both polarimeters have similar design: incoming light passes through a modulator (superachromatic half-wavelength plate, HWP),  it is then split by the plane-parallel calcite plate into two orthogonally polarized rays (ordinary and extraordinary) and further is reflected by two dichroic beam splitters to produce o- and e-images on three charge-coupled devices (CCDs) simultaneously in the $B, V, R$ bands.  
This design optically eliminates the sky polarization at the registration stage (even if it is variable), reaching up to $10^{-5}$ polarization precision \cite{PBB14,Berdyugin19,Piirola21}. 
The instrumental polarization is below $10^{-4}$ level and can be accounted for using zero-polarization standards\cite{Piirola21} but is negligible for our study. 
The zero point of the PA was determined by observing high polarization standards HD 161056 and HD 204827\cite{Piirola21}.

For each independent measurement of linear polarization, four sequential images with the HWP rotated by $22\fdg5$ in each step are obtained.
Each measurement cycle consists of 16 images recorded simultaneously by three CCDs for one full rotation of the HWP modulator (0$\degr$--360$\degr$) giving four independent measurements of the normalized Stokes parameters $(q,u)$ in the $B$, $V$, and $R$ bands.
This algorithm helps eliminate effects arising from dust particles on the retarder, nonparallelism of rotating components, etc.
The images produced by DIPol-UF require standard CCD calibrations\cite{Berdyugin19}.

The average normalized Stokes parameters $(q, u)$ are then  obtained from 16 to 40 individual measurements using a $2\sigma$ weighting algorithm\cite{Kosenkov17,Piirola21}.
The typical duration of an observational window where the data were averaged was one hour. 
The polarization produced by the interstellar medium (ISM) has been estimated from numerous (more than 400) observations of sample of field stars (stars \#2, 3, 6, 7, 9 from fig.~9 of ref. \cite{Veledina19}), which are close in distance to the target as indicated by their parallaxes\cite{Gaia18}.
These normalized Stokes parameters $(q_{\rm ISM}, u_{\rm ISM})$ were subtracted from the measured values of the normalized Stokes parameters to obtain the intrinsic polarization information of the source $(q_{\rm intr}, u_{\rm intr})$. 
We then define the complex linear polarization quantity as ${\cal P}=q_{\rm intr} +i u_{\rm intr}$. 
Intrinsic polarization degree (PD) $P$ and polarization angle (PA) $\theta$ are then obtained from the formulae 
\begin{equation} 
 P = |{\cal P}|  = \sqrt{q_{\rm intr}^2+u_{\rm intr}^2}, \quad 
\theta= \frac{1}{2} \arg ({\cal P})  .
\end{equation} 
Because our PD measurements typically have significance 5--10$\sigma$, the bias in the measured PD is negligible and the uncertainty in PD $\Delta P$ is the same as the uncertainty on individual Stokes parameters. 
The uncertainty of the PA is estimated as  $\Delta \theta =  \Delta P/(2P)$ \cite{Serkowski62}.  
The intrinsic PD and PA computed following this procedure are reported in Table~\ref{tab:dipol}.

Polarization observed during quiescence shows no clear dependence on the orbital phase (Fig.~\textbf{2}), apart from some spread of the points in close orbital phases. 
The PD has a blue spectrum depending on frequency $\nu$  as  $\propto\nu^{3}$, which is inconsistent with the red spectra expected from a jet or accretion disk.
The PA is very stable in the $B$-band, where the PD is the highest and the relative uncertainty is smallest.  
Therefore, we computed the average PA in that band as a inverse-variance weighted mean of individual PAs. 
Since the deviations of individual measurements exceed the measurement uncertainties, the  standard error of the weighted mean was corrected for overdispersion by the square root of the reduced $\chi^2$, resulting in $\langle \mbox{PA}\rangle=-19\fdg7\pm1\fdg2$. 
We also computed the mean and its uncertainty using a bootstrap (sampling with replacement) method\cite{Efron79} arriving at an identical value.  
The mean PA in the $V$- and $R$-bands have similar values, but larger uncertainties (Table~\ref{tab:dipol}). 

\renewcommand{\thefigure}{{\bf S1}}
\begin{figure}
\centering
\includegraphics[width=0.45\columnwidth]{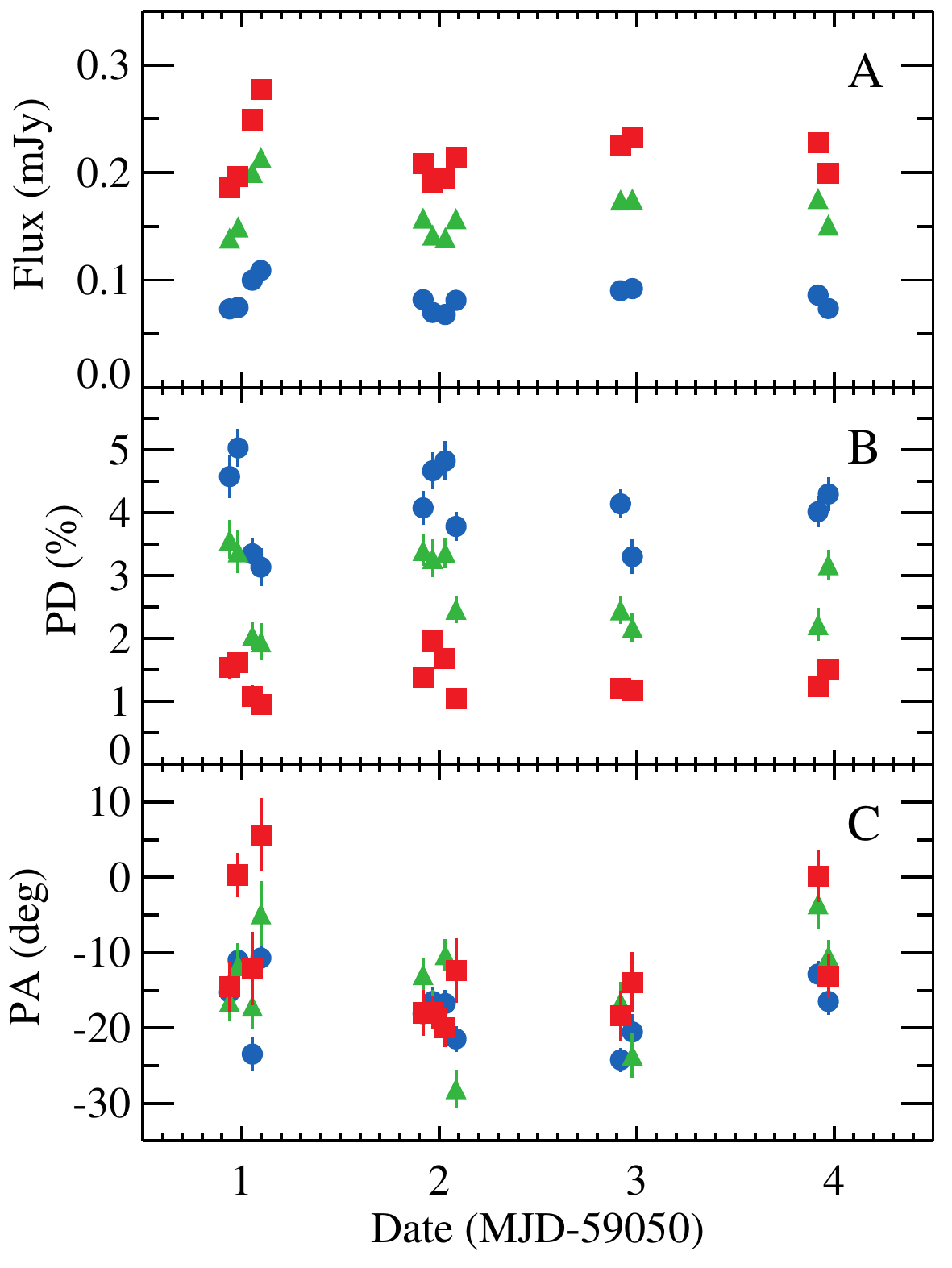}
\includegraphics[width=0.45\columnwidth]{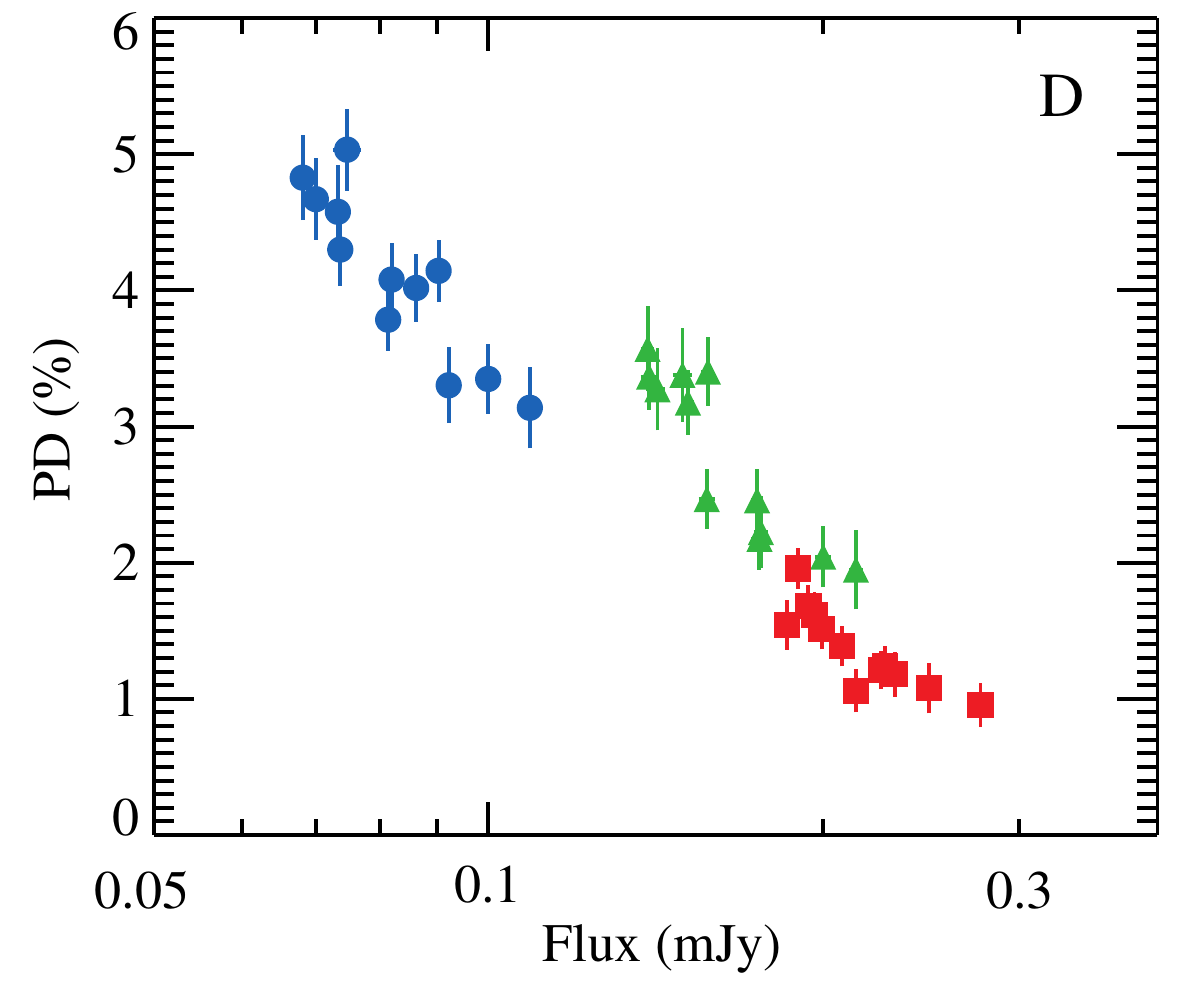}
\caption{\textbf{Variation of flux and polarization.} 
\textbf{(A)} The dependence of the observed flux (not corrected for reddening), \textbf{(B)} intrinsic source PD and \textbf{(C)} PA on time 
 during the July 2020 observations of \maxi\ with DIPol-UF in three filters: $B$ (blue circles), $V$ (green triangles), and $R$ (red squares).  
\textbf{(D)} Dependence of the PD on flux in the same three filters. An anticorrelation is evident. 
\label{fig:flux_PD_PA}}
\end{figure}

\renewcommand{\thetable}{{\bf S1}}
\begin{table*}
 \centering
\caption{\textbf{Intrinsic polarization of \maxi.} 
The intrinsic values of the PD and PA derived from observations of \maxi\ during the outburst with DIPol-2\cite{Veledina19,Kosenkov20} and during quiescence with DIPol-UF at NOT. Uncertainties are 1$\sigma$. The Stokes parameters have been corrected for the interstellar polarization. The dates for observations during quiescence correspond to the middle of observing windows of a typical duration of one hour. 
\label{tab:dipol}}
\begin{footnotesize}
\begin{tabular}{crrcrrcrr}
    \hline
Date & 
\multicolumn{2}{c}{$B$} &  & 
\multicolumn{2}{c}{$V$} &  &  \multicolumn{2}{c}{$R$} \\ 
\cline{2-3} 
\cline{5-6}
\cline{8-9}
 (MJD)  & PD  (\%) & PA (deg) & & PD  (\%) & PA (deg) & & PD  (\%) & PA (deg)    \\
        \hline
 \multicolumn{9}{c}{Outburst}    \\
 58195--58222 &    $0.28\pm0.01$ &    $9.2 \pm1.0$ & &   $0.36\pm0.01$ &    $22.9\pm1.0$ & &    $0.30\pm    0.01$ &   $29.0\pm0.9$ \\
58223--58234 &    $0.34\pm0.02$ &    $8.8\pm1.4$ &&    $0.51\pm0.02$ &    $23.4\pm1.4$ &&    $0.53 \pm0.02$ &   $23.9 \pm1.1$ \\
58312--58344   &    $0.16 \pm0.01$ &   $-15.8\pm1.6$ & &   $0.15 \pm0.01$ &    $13.4\pm2.3$ &&    $0.02 \pm   0.01$ &   $39.1\pm11.0$ \\
58406--58428  &    $0.06\pm0.04$ &    $-3.0\pm15.4$ & &   $0.13\pm0.06$ &     $2.8\pm12.4$ &&    $0.09\pm0.04$ &   $44.8 \pm12.5$ \\
\multicolumn{9}{c}{Quiescence}    \\ 
  58686.93 & $3.86\pm0.29$ & $-23.8\pm 2.1$& & $1.81\pm0.24$ & $-23.8\pm 3.8$& & $0.70\pm0.14$ & $-24.1\pm 5.7$\\
  58688.01 & $2.35\pm0.38$ & $-25.9\pm 4.7$& & $1.09\pm0.25$ & $  8.0\pm 6.5$& & $1.70\pm0.10$ & $-42.4\pm 1.7$\\
  58961.15 & $1.47\pm0.28$ & $-26.8\pm 5.4$& & $0.52\pm0.24$ & $ -4.2\pm13.3$& & $0.63\pm0.19$ & $  8.5\pm 8.7$\\
  58964.16 & $4.31\pm0.28$ & $-18.5\pm 1.8$& & $3.10\pm0.29$ & $-17.5\pm 2.7$& & $1.05\pm0.18$ & $ -2.0\pm 4.8$\\
  58965.11 & $2.81\pm0.24$ & $-23.5\pm 2.5$& & $2.12\pm0.21$ & $-35.1\pm 2.9$& & $0.84\pm0.18$ & $-26.3\pm 6.2$\\
  58967.12 & $4.53\pm0.28$ & $-24.1\pm 1.8$& & $2.09\pm0.39$ & $-11.0\pm 5.3$& & $1.09\pm0.31$ & $-27.0\pm 8.3$\\
  59050.94 & $4.58\pm0.34$ & $-15.3\pm 2.1$& & $3.57\pm0.31$ & $-16.5\pm 2.5$& & $1.54\pm0.18$ & $-14.5\pm 3.4$\\
  59050.98 & $5.03\pm0.30$ & $-11.0\pm 1.7$& & $3.38\pm0.34$ & $-11.7\pm 2.9$& & $1.62\pm0.16$ & $  0.3\pm 2.9$\\
  59051.05 & $3.35\pm0.26$ & $-23.5\pm 2.2$& & $2.04\pm0.23$ & $-17.1\pm 3.2$& & $1.08\pm0.18$ & $-12.1\pm 4.9$\\
  59051.10 & $3.14\pm0.30$ & $-10.7\pm 2.7$& & $1.95\pm0.29$ & $ -4.8\pm 4.3$& & $0.95\pm0.16$ & $  5.6\pm 4.9$\\
  59051.92 & $4.08\pm0.27$ & $-18.1\pm 1.9$& & $3.40\pm0.25$ & $-12.9\pm 2.1$& & $1.39\pm0.15$ & $-18.0\pm 3.1$\\
  59051.97 & $4.67\pm0.30$ & $-16.5\pm 1.8$& & $3.27\pm0.30$ & $-17.5\pm 2.6$& & $1.96\pm0.15$ & $-17.9\pm 2.2$\\
  59052.03 & $4.83\pm0.31$ & $-16.7\pm 1.9$& & $3.37\pm0.24$ & $-10.3\pm 2.1$& & $1.68\pm0.15$ & $-20.0\pm 2.6$\\
  59052.09 & $3.78\pm0.23$ & $-21.5\pm 1.7$& & $2.47\pm0.22$ & $-28.0\pm 2.5$& & $1.06\pm0.16$ & $-12.4\pm 4.3$\\
  59052.92 & $4.14\pm0.23$ & $-24.2\pm 1.6$& & $2.46\pm0.23$ & $-16.6\pm 2.7$& & $1.21\pm0.14$ & $-18.4\pm 3.3$\\
  59052.98 & $3.30\pm0.28$ & $-20.5\pm 2.4$& & $2.18\pm0.23$ & $-23.6\pm 3.0$& & $1.18\pm0.17$ & $-14.0\pm 4.0$\\
  59053.92 & $4.02\pm0.25$ & $-12.8\pm 1.8$& & $2.22\pm0.26$ & $ -3.5\pm 3.4$& & $1.24\pm0.15$ & $  0.2\pm 3.4$\\
  59053.97 & $4.30\pm0.27$ & $-16.5\pm 1.8$& & $3.18\pm0.24$ & $-10.4\pm 2.1$& & $1.51\pm0.15$ & $-13.1\pm 2.9$\\
 59400.99 & $1.62\pm0.12$ & $-30.5\pm 2.0$& & $0.77\pm0.10$ & $-17.1\pm 3.9$& & $0.41\pm0.07$ & $  6.4\pm 5.1$\\
 59401.94 & $2.86\pm0.21$ & $-25.4\pm 2.1$& & $2.04\pm0.20$ & $-29.9\pm 2.8$& & $0.36\pm0.14$ & $ -6.1\pm10.7$\\ 
59051--59054 & $4.00\pm0.19$ &  $-17.5\pm 1.3$ & & $2.65\pm0.19$ & $-14.7\pm 1.9$  & &  $1.32\pm0.08$  & $-12.8\pm 2.3$ \\  
58686--59402 & $3.18\pm0.22$ & $-19.7\pm 1.2$ & & $1.89\pm0.19$ & $-16.8\pm1.9$ & &  $0.94\pm0.11$  & $-18.6\pm 3.3$ \\   
 \multicolumn{9}{c}{Interstellar polarization}    \\ 
58195--59054  & $0.81\pm0.03$ & $64.0\pm1.1$ & & 
$0.71\pm0.03$ & $68.4\pm1.2$ & &  
$0.60\pm0.02$ & $64.4\pm0.8$  \\ 
\hline  
\end{tabular}
\end{footnotesize}
\end{table*}

\renewcommand{\thetable}{{\bf S2}}
\begin{table*}
 \centering
\caption{\textbf{Observed flux  of \maxi.} 
Mean fluxes and their $1\sigma$ uncertainties derived from observations of \maxi\ during quiescence in July 2020 with DIPol-UF at NOT.    
\label{tab:flux_not}}  
\begin{footnotesize}
\begin{tabular}{cccc}
    \hline
Date  & \multicolumn{3}{c}{Flux (mJy)} \\ 
\cline{2-4}
(MJD) &  $B$ &  $V$ & $R$  \\
        \hline
    59050.94 & $0.0732\pm0.0014$ & $0.1391\pm0.0019$ & $0.1858\pm0.0020$ \\
    59050.98 & $0.0747\pm0.0022$ & $0.1495\pm0.0030$ & $0.1966\pm0.0029$ \\
    59051.05 & $0.1000\pm0.0022$ & $0.2002\pm0.0034$ & $0.2491\pm0.0035$ \\
    59051.10 & $0.1090\pm0.0018$ & $0.2142\pm0.0030$ & $0.2772\pm0.0028$ \\
    59051.92 & $0.0819\pm0.0015$ & $0.1577\pm0.0021$ & $0.2082\pm0.0023$ \\
    59051.97 & $0.0700\pm0.0014$ & $0.1420\pm0.0022$ & $0.1901\pm0.0022$ \\
    59052.03 & $0.0681\pm0.0015$ & $0.1395\pm0.0024$ & $0.1941\pm0.0025$ \\
    59052.09 & $0.0813\pm0.0018$ & $0.1572\pm0.0026$ & $0.2142\pm0.0027$ \\
    59052.92 & $0.0902\pm0.0015$ & $0.1745\pm0.0023$ & $0.2255\pm0.0023$ \\
    59052.98 & $0.0922\pm0.0021$ & $0.1754\pm0.0031$ & $0.2323\pm0.0031$ \\
    59053.92 & $0.0861\pm0.0017$ & $0.1759\pm0.0027$ & $0.2275\pm0.0028$ \\
    59053.97 & $0.0736\pm0.0013$ & $0.1512\pm0.0020$ & $0.1995\pm0.0020$ \\
\hline  
\end{tabular}
\end{footnotesize}
\end{table*}

Using the images from DIPol-UF we also measured photometry of \maxi\ relative to two nearby stars (\#1 and 2 from table 2 in ref. \cite{Veledina19}).   
Absolute photometry during July 2020 observations was obtained using simultaneous observations by the Liverpool Telescope (LT), see Fig.~\ref{fig:flux_PD_PA}A and Table~\ref{tab:flux_not}. 
The flux of \maxi\ in all bands is anti-correlated with the PD  (Fig.~\ref{fig:flux_PD_PA}D). 
This anti-correlation together with the spectral dependence of the PD can be interpreted as an interplay between two components: one, polarized, with contribution growing towards blue and another, unpolarized or, possibly, weakly polarized, variable component dominating the flux in the red. 
Following the spectral decomposition (Fig.~\textbf{2}), the UV component produces only 20\% of the total flux in the $R$ band.
This makes PA measurements of the UV component less reliable in this band.
Stronger variations of the PA in $R$ may be caused by the presence of a red and polarized at a level of 0.1--0.3\%  component, whose PA is different from that of the UV component.
Such red and polarized component, with the PA consistent with the jet position angle, has been observed during the outburst (Table~\ref{tab:dipol}).

\subsection*{Liverpool Telescope observations}
\label{sec:LT}

Photometric observations in the optical band were performed using the optical imaging component of the Infrared-Optical suite of instruments (IO:O) with the Sloan  Digital Sky Survey (SDSS) $i'r'z'$, Bessel $V$ and $B$ filters on the LT\cite{Steele04} on La Palma, Spain.
The object was observed during four nights from  2020 July 20 to 24, i.e. Modified Julian Dates (MJD) 59050--59054. 
LT image reduction was provided by the basic IO:O pipeline\cite{lt_ioo}.
As the photometric standards, we used four stars with known Panoramic Survey Telescope and Rapid Response System (PS1) point spread function magnitudes. 
To convert PS1 magnitudes to SDSS and Johnson-Cousins magnitudes we used equation (6) and table 6 from ref.\cite{Tonry12}.
To obtain fluxes of the object from its magnitudes we used standard zero-points for SDSS and Johnson-Cousins systems\cite{Fukugita96,Bessell79}.
The fluxes and their uncertainties are presented in Table~\ref{tab:obs_LT}.

\renewcommand{\thetable}{{\bf S3}}
\begin{table*}
 \centering
\caption{\textbf{Log of observations with the Liverpool Telescope.} 
The observed fluxes of \maxi\ and their $1\sigma$ uncertainties in various filters.
\label{tab:obs_LT} }
\begin{footnotesize}
\begin{tabular}{ccrccccrc} 
    \hline
Date   & Filter   &  Exposure & Flux   && Date   & Filter   &  Exposure & Flux   \\ 
(MJD)  &          &  (s)      & (mJy)   && (MJD)  &          &  (s)      & (mJy)  \\
        \hline
59050.92075 & $i$ &   100 & 0.2700$\pm$0.0014 && 59052.92842 & $z$ &   100 & 0.3944$\pm$0.0033\\
59050.92213 & $i$ &   100 & 0.2468$\pm$0.0015 && 59052.92980 & $z$ &   100 & 0.4037$\pm$0.0033\\
59050.92350 & $i$ &   100 & 0.2482$\pm$0.0015&&59052.93193 & $V$ &   200 & 0.1605$\pm$0.0011\\
59050.92505 & $z$ &   100 & 0.3990$\pm$0.0028&&59052.93463 & $B$ &   200 & 0.0974$\pm$0.0007\\
59050.92642 & $z$ &   100 & 0.4093$\pm$0.0029&&59052.93637 & $r$ &    40 & 0.3063$\pm$0.0018\\
59050.92780 & $z$ &   100 & 0.3733$\pm$0.0029&&59052.93705 & $r$ &    40 & 0.2142$\pm$0.0018\\
59050.93052 & $V$ &   300 & 0.1440$\pm$0.0009&&59053.92120 & $i$ &   100 & 0.3071$\pm$0.0017\\
59051.01279 & $i$ &   100 & 0.3321$\pm$0.0014&&59053.92257 & $i$ &   100 & 0.3600$\pm$0.0018\\
59051.01417 & $i$ &   100 & 0.3075$\pm$0.0014&&59053.92395 & $i$ &   100 & 0.3450$\pm$0.0018\\
59051.01554 & $i$ &   100 & 0.3164$\pm$0.0014&&59053.92550 & $z$ &   100 & 0.3917$\pm$0.0033\\
59051.01709 & $z$ &   100 & 0.4208$\pm$0.0027&&59053.92687 & $z$ &   100 & 0.3598$\pm$0.0033\\
59051.01846 & $z$ &   100 & 0.5048$\pm$0.0028&&59053.92825 & $z$ &   100 & 0.3736$\pm$0.0034\\
59051.01984 & $z$ &   100 & 0.4703$\pm$0.0027&&59053.93038 & $V$ &   200 & 0.1369$\pm$0.0011\\
59051.02254 & $B$ &   300 & 0.1263$\pm$0.0005&&59053.93308 & $B$ &   200 & 0.0762$\pm$0.0007\\
59051.09342 & $i$ &   100 & 0.3476$\pm$0.0016&&59053.93482 & $r$ &    40 & 0.2281$\pm$0.0019\\
59051.09479 & $i$ &   100 & 0.3489$\pm$0.0016&&59053.93550 & $r$ &    40 & 0.2174$\pm$0.0019\\
59051.09617 & $i$ &   100 & 0.3809$\pm$0.0015&&59053.95474 & $i$ &   100 & 0.2752$\pm$0.0016\\
59051.09771 & $z$ &   100 & 0.4877$\pm$0.0032&&59053.95612 & $i$ &   100 & 0.2927$\pm$0.0016\\
59051.09909 & $z$ &   100 & 0.4963$\pm$0.0033&&59053.95750 & $i$ &   100 & 0.2633$\pm$0.0017\\
59051.10047 & $z$ &   100 & 0.4706$\pm$0.0033&&59053.95904 & $z$ &   100 & 0.3587$\pm$0.0032\\
59051.10168 & $r$ &    40 & 0.2339$\pm$0.0019&&59053.96042 & $z$ &   100 & 0.3824$\pm$0.0030\\
59051.10236 & $r$ &    40 & 0.2592$\pm$0.0018&&59053.96180 & $z$ &   100 & 0.4097$\pm$0.0033\\
59051.92206 & $i$ &   100 & 0.3003$\pm$0.0015&&59053.96393 & $V$ &   200 & 0.1467$\pm$0.0010\\
59051.92343 & $i$ &   100 & 0.2982$\pm$0.0015&&59053.96663 & $B$ &   200 & 0.0799$\pm$0.0007\\
59051.92481 & $i$ &   100 & 0.2860$\pm$0.0015&&59053.97116 & $i$ &   100 & 0.3032$\pm$0.0042\\
59052.01060 & $i$ &   100 & 0.2838$\pm$0.0016&&59053.97254 & $i$ &   100 & 0.3390$\pm$0.0016\\
59052.01197 & $i$ &   100 & 0.2693$\pm$0.0015&&59053.97391 & $i$ &   100 & 0.3308$\pm$0.0016\\
59052.01335 & $i$ &   100 & 0.2501$\pm$0.0016&&59053.97546 & $z$ &   100 & 0.4278$\pm$0.0032\\
59052.09336 & $i$ &   100 & 0.2979$\pm$0.0015&&59053.97684 & $z$ &   100 & 0.4017$\pm$0.0032\\
59052.09474 & $i$ &   100 & 0.2800$\pm$0.0016&&59053.97821 & $z$ &   100 & 0.4193$\pm$0.0031\\
59052.09611 & $i$ &   100 & 0.2974$\pm$0.0015&&59053.98035 & $V$ &   200 & 0.1535$\pm$0.0010\\
59052.92275 & $i$ &   100 & 0.3161$\pm$0.0017&&59053.98305 & $B$ &   200 & 0.0601$\pm$0.0007\\
59052.92412 & $i$ &   100 & 0.3058$\pm$0.0016&&59053.98479 & $r$ &    40 & 0.2427$\pm$0.0017\\
59052.92550 & $i$ &   100 & 0.3048$\pm$0.0017&&59053.98546 & $r$ &    40 & 0.2100$\pm$0.0017\\ 
59052.92705 & $z$ &   100 & 0.3686$\pm$0.0034&& &   &     &    \\ 
       \hline  
\end{tabular}
\end{footnotesize}
\end{table*}

\subsection*{\textit{Swift}/UVOT observations}

The \textit{Neil Gehrels Swift Observatory} (\textit{Swift})\cite{Gehrels04}  observed \maxi\ in the low state with the X-ray Telescope (XRT)  and Ultraviolet/Optical Telescope (UVOT) instruments (partly quasi-simultaneously with DIPol-UF/NOT) between 2020 July 20 -- September 6.
All the XRT data were taken in the photon counting mode with about 9 ks of total exposure. 
An averaged spectrum was extracted using the online tools\cite{EBP2009} provided by the UK Swift Science Data Centre. 
The spectrum was rebinned to have at least 1 count per energy channel in order to apply $W$-statistic\cite{Wachter79,xspec_statistics}.
The data were fitted with a model consisting of a power law modified by the interstellar absorption {\sc tbabs*powerlaw} in the 0.5--10~keV band using the {\sc xspec} package\cite{Arn96}. 
Fixing the hydrogen column density to $N_{\rm H}=1.6\times10^{21}$\,cm$^{-2}$\cite{Bharali19} we obtained a power-law photon index $\Gamma=1.62\pm0.24$ with $W$-statistics of 42.7 for 46 degrees of freedom. 
We find flux corrected for absorption of $(3.3\pm 0.7) \times 10^{-13}\,\textrm{erg}\,\textrm{cm}^{-2}\,\textrm{s}^{-1}$ in the 0.5--10 keV energy range.

The UVOT data were processed with the \textsc{uvotproduct} tool\cite{Breeveld11,BBR2016,swift_uvot}.
We used a $5\arcsec$ source aperture around the known position of \maxi\cite{Gaia18}, and a circular source-free $10\arcsec$ background aperture about $40\arcsec$ east of it.
After comparing the magnitudes obtained between July 20 -- September 6 to other quiescence data taken when the x-ray counting rate was low, we found that during this period \maxi\ was roughly 0.5--1 magnitude fainter in all the UV filters than in the data taken in 2019 or earlier in 2020.
The same trend was observed in the $V$-band and, to a lesser extent, in the $R$-band.
Therefore, we decided to stack all the 2020 July 20--September 6 UVOT data together to obtain time averaged fluxes in the \textit{u}, \textit{uvw1}, \textit{uvm2} and \textit{uvw2} filters for the spectral energy distribution (SED).
Conversion of the background corrected count rates obtained from the \textsc{uvotproduct} to the fluxes is non-trivial, as it depends on the assumed spectral shape\cite{BBR2016}. 
Therefore, for reliable estimate of the fluxes we performed direct spectral fitting with a broad-band SED model using the {\sc xspec} package (Fig.~\textbf{2}).

\subsection*{Decomposition of the broadband spectrum}
\label{sec:SED}

The broad-band average spectrum obtained with the LT and \textit{Swift}/UVOT is shown in Fig.~\textbf{2}.  
The spectrum was corrected for reddening using model of ref. \cite{Cardelli89}  with extinction $A_V=0.89$ (corresponding to the color excess $E(B-V)=0.29$) obtained from the hydrogen column density $N_{\rm H}=(0.16\pm0.02)\times10^{22}\mbox{cm}^{-2}$\cite{Bharali19}  using transformation $A_V=5.59\, N_{\rm H}/10^{22}\,\mbox{cm}^{-2}$\cite{Predehl95}.
The optical/infrared SED obtained  with the LT is very red and can be described by a joined contribution of companion star and  multicolor accretion disk, but the UVOT data show a UV excess. 
The excess is also seen in the non-dereddened data and is not an artefact of dereddening procedure. 
A similar excess has been detected previously in quiescent-state observations of the black hole x-ray binary A~0620$-$00\cite{Froning11}, but its nature remained a mystery.

The blue PD spectrum differs from the red (even after correcting for the reddening) spectrum of the infrared to UV continuum. 
This rules out accretion disk as the source of polarized light. 
We infer that the component producing the UV excess is responsible for polarization. 
To extract the average shape of the polarized component, we calculated the polarized fluxes as
\begin{equation}
(PF)_{k} = P_{k}F_{k},
\end{equation}
where $k$ is the index corresponding to one of the $BVR$-bands, $P_{k}$ is the average PD and $F_{k}$ is the average flux in this band. 
If PD of the polarized component is wavelength-independent, then the computed $PF$ directly replicates its total spectral shape, with the normalization being  smaller by the polarization fraction.
The average PD was computed from the average Stokes parameters $(q_{\rm intr},u_{\rm intr})$ obtained in 12 individual measurements in July 2020 during quiescence with DIPol-UF (Table~\ref{tab:dipol}). 
Because absolute photometry with DIPol-UF is not very reliable, the average flux was obtained from the LT observations. 
The averages were computed as a inverse-variance weighted mean of individual measurements and their standard errors were corrected for overdispersion by the square root of the reduced $\chi^2$.

Published spectroscopy obtained during quiescence indicates a K-type spectral type for the companion star, which contributed about 20\% to the flux in the $r$-band\cite{Torres19,Torres20}. 
The average observed $r$-band flux during the spectroscopy was about 0.33 mJy, higher than the average flux of 0.24~mJy in our LT observations.
Thus, the contribution of the companion star to the $r$-band flux rose to 27\%.
The radius of the companion star, which fills its Roche lobe, can be established from the measured orbital period $P_{\rm orb}=0.68549\pm0.00001$~d, projected rotational velocity $v_{\rm rot}\sin i_{\rm orb}=84\pm5$~km~s$^{-1}$, and orbital inclination $i_{\rm orb}=73\degr \pm6\degr$\cite{Torres20} 
\begin{equation} 
R_{\rm c}  = \frac{P_{\rm orb}(v_{\rm rot}\sin i_{\rm orb})}{2\pi \sin i_{\rm orb}} 
= (1.19\pm0.08) R_\odot = (8.3\pm0.6)\times 10^{10}\ \mbox{cm}.
\end{equation}
A moderately evolved late-type star of spectral class K7 can satisfy the constraint on radius and contribute about 27\% to the $r$-band flux.
A star with higher temperature overpredicts the companion star contribution in the $r$-band.
For the fitting purposes, we approximate the stellar spectrum\cite{Pickles98} by the blackbody with $T_*=4000$~K and $R_*=R_{\rm c}=8.3\times 10^{10}$~cm (see Fig.~\textbf{2}).  

Using {\sc xspec} v.12.11\cite{Arn96},  we modelled the total SED from LT and \textit{Swift}/UVOT and, jointly, the three points of the $PF$. 
The total spectrum is described by the model {\sc redden} ({\sc bbodyrad}$_1$+{\sc diskbb}+{\sc bbodyrad}$_2$). 
The {\sc redden} model describes interstellar extinction \cite{Cardelli89}.
Component {\sc bbodyrad}$_1$  corresponds to the spectrum of the stellar companion, modeled as a blackbody of fixed temperature $T_*$ and radius $R_*$. 
The {\sc diskbb} corresponds to the multicolor accretion disk. It has two free parameters: normalization, which is related to the inner radius $R_{\rm in}$, and temperature at that radius $T_{\rm in}$. 
Component {\sc bbodyrad}$_2$  corresponds to the UV excess that is modeled by a blackbody with two free parameters: temperature $T_{\rm bb}$ and radius $R_{\rm bb}$. 
The $PF$ is modeled by the reddened second blackbody multiplied by the polarization fraction $P_{\rm UV}$.
Blackbody and disk normalizations were converted to radii assuming the distance to the source of 2.96~kpc as determined by the radio parallax\cite{Atri20} and disk inclination $i_{\rm orb}=73\degr$\cite{Torres20}. 

The best-fitting model parameters with the corresponding uncertainties are listed in Table~\ref{tab:sed}.
We find the accretion disk  temperature of about 6200~K, which is very close to that expected for the disk in quiescence\cite{Frank02}. 
The inner radius of the disk $R_{\rm in}\approx 6\times10^{10}$~cm, which is 30\% smaller than the estimate of the circularization radius for the measured $P_{\rm orb}$ and companion-to-black hole mass ratio $q=0.072\pm0.012$\cite{Torres20}.
The outer radius has to be at least 2--3 times larger to avoid underpredicting the red part of the spectrum.
This is consistent with the expectation that the accretion disk size does not exceed the tidal radius of about $2.4\times 10^{11}$~cm\cite{Frank02}.

\renewcommand{\thetable}{{\bf S4}}
\begin{table}
\centering
\caption{\textbf{Best-fitting parameters of the SED model.} 
Distance of 2.96 kpc\cite{Atri20} and inclination $i_{\rm orb}=73\degr$\cite{Torres20} are assumed. 
The temperature and the radius of the blackbody approximating the SED of the stellar companion were fixed.
  \label{tab:sed}}
    \begin{tabular}{l l  l} 
    \hline
        Parameter & Value  & Units\\
        \hline
        $T_{\rm in}$ & $6200^{+1400}_{-1100}$ & K  \\
        $R_{\rm in}$ & $(5.6^{+2.7}_{-1.8})\times10^{10}$ &  cm \\
        $T_{\rm bb}$ & $14900^{+2300}_{-1400}$  &  K \\ 
        $R_{\rm bb}$ & $(8.9^{+2.3}_{-2.1})\times10^{9}$ & cm  \\
         $T_*$ & $4000$ &   K \\ 
        $R_*$ & $8.3\times10^{10}$  &   cm  \\   
        $P_{\rm UV}$    & $0.055^{+0.023}_{-0.011}$   &  \\
    \hline
    \end{tabular}
\end{table}

The temperature of the additional blackbody is $T_{\rm bb}\sim15,000$~K and the characteristic size $R_{\rm bb}\sim9\times10^{9}$~cm.
This reproduces the shape of the polarized flux SED and the UV excess.
The PD of this component is $P_{\rm UV}=0.055^{+0.023}_{-0.011}$, corresponding to 5--8\% intrinsic PD (on average) of the polarized component. 
The broad-band SED and the best-fitting spectral components are shown in Fig.~\textbf{2}.
The computed values of $PF$ were divided by the constant $P_{\rm UV}$ to demonstrate the fit with the UV excess component.

For lower and higher values of the color excess of 0.25 and 0.325, corresponding to the 1$\sigma$ uncertainties in $N_{\rm H}$, the best-fitting spectral parameters change.
The lower $E(B-V)$ leads to the 8\% smaller disk temperature $T_{\rm in}$ and 13\% larger inner radius $R_{\rm in}$, 8\% smaller blackbody temperature $T_{\rm bb}$  and 3\% larger radius $R_{\rm bb}$. 
For the higher $E(B-V)$, the effect is opposite: $T_{\rm in}$ is 7\% larger,  $R_{\rm in}$ is  9\% smaller, $T_{\rm bb}$ is 7\% larger and $R_{\rm bb}$ is smaller by 3\%.  
This uncertainty does not affect any of our conclusions.

\subsection*{The source of polarized light and the nature of the spectral components}
\label{sec:nature}

In this section we seek to answer two questions: what is the nature of the UV excess and what is the source of polarized radiation?
The accretion disk itself cannot be the source of the UV emission, because in quiescence its temperature is lower as we see from the red optical spectrum.
Another possibility is a hotspot (or hot line), the place where the accretion stream hits the disk. 
This component could be responsible for the UV excess. 
However, whether it can also be the source of polarization is questionable. 
The high temperature of this component implies that the matter is ionized. 
The PD from the optically thick electron-scattering dominated atmosphere\cite{Cha60,Sob63} depends on the cosine of the viewing angle $\mu=\cos i_{\rm orb}$ approximately as $11.7\% \times(1-\mu)/(1+3.582\mu)$\cite{Viironen04}. 
For inclination angles permitted by the absence of eclipses\cite{Torres20}, $i_{\rm orb}<81\degr$, the expected PD is $\lesssim6$\%. 
A high implied PD, $P_{\rm UV}\approx$5--8\%, is barely consistent with that. 

On the other hand, high PD can be produced by synchrotron radiation in the ordered magnetic field.
The blue spectrum indicates that we might see the optically thick part of this radiation, with the possible transition to optically thin part at $\nu\gtrsim10^{15}$~Hz.
Self-absorption becomes important at the turn-over frequency\cite{RL79,WZ01,VPV13},
$\displaystyle \nu_{\rm t}\approx3\times10^{15} B_{6}^{\frac{p+2}{p+4}}(\tau_{\rm t}\gamma_{\rm t}^p)^{\frac{2}{p+4}}$, where $B_{6}$ is the magnetic field in units of $10^6$~G, $\tau_{\rm t}$ is the Thomson optical depth of electrons emitting at the turn-over frequency, $\gamma_{\rm t}$ is their Lorentz factor and $p$ is the index of the power-law distribution of electron number density on the Lorentz factor $\gamma$, $dn_{\rm e}/d\gamma\propto \gamma^{-p}$. 
For the observed $\nu_{\rm t}\approx10^{15}$~Hz, this requires highly opaque source with $\tau_{\rm t}\gamma_{\rm t}^p\sim 1$ and/or high magnetic field with $B_{6}\sim1$, which would be expected in the bright hard state, but inconsistent with the relevant values for the source in (near-)quiescence, when both optical depth and the magnetic field drop by 2 to 4 orders of magnitude\cite{YN14}.  
The highest theoretically possible PD below the self-absorption frequency\cite{Ginzburg69}  of $P_{\max}=3/(6p+13)\sim 10\%$ requires highly ordered -- and constant -- magnetic field during two years, July 2019--July 2021. 
Hence, we find the synchrotron radiation to be an implausible source of the observed polarized flux.

Substantial polarization could instead be produced by electron scattering of radiation in an optically thin slab if the seed photons are injected along the slab plane\cite{ST85}.
The PD$=(1-\mu^2)/(3-\mu^2)$ reaches 33\% edge-on and is a weak function of orbital inclination (for $i_{\rm orb}\gtrsim 66\degr$, i.e. $\mu\lesssim 0.41$, the PD is larger than $\sim$30\%).
Scattering in the Thomson regime means that the scattered radiation does not gain a systematic shift in energy with respect to the incident continuum, and hence the peak of the spectrum of the polarized component directly probes the characteristic energy and spectral shape of the incident radiation. 
Hence, the source of the incident light should have narrow, blackbody-like shape, with characteristic  temperature $T\sim15,000$~K.
Because this temperature is much higher than the disk inner temperature, and the polarization angle is independent of orbital phase, the location and physical properties of such a component are unclear.

\renewcommand{\thefigure}{{\bf S2}} 
\begin{figure} 
\centering
\includegraphics[width=0.5\columnwidth]{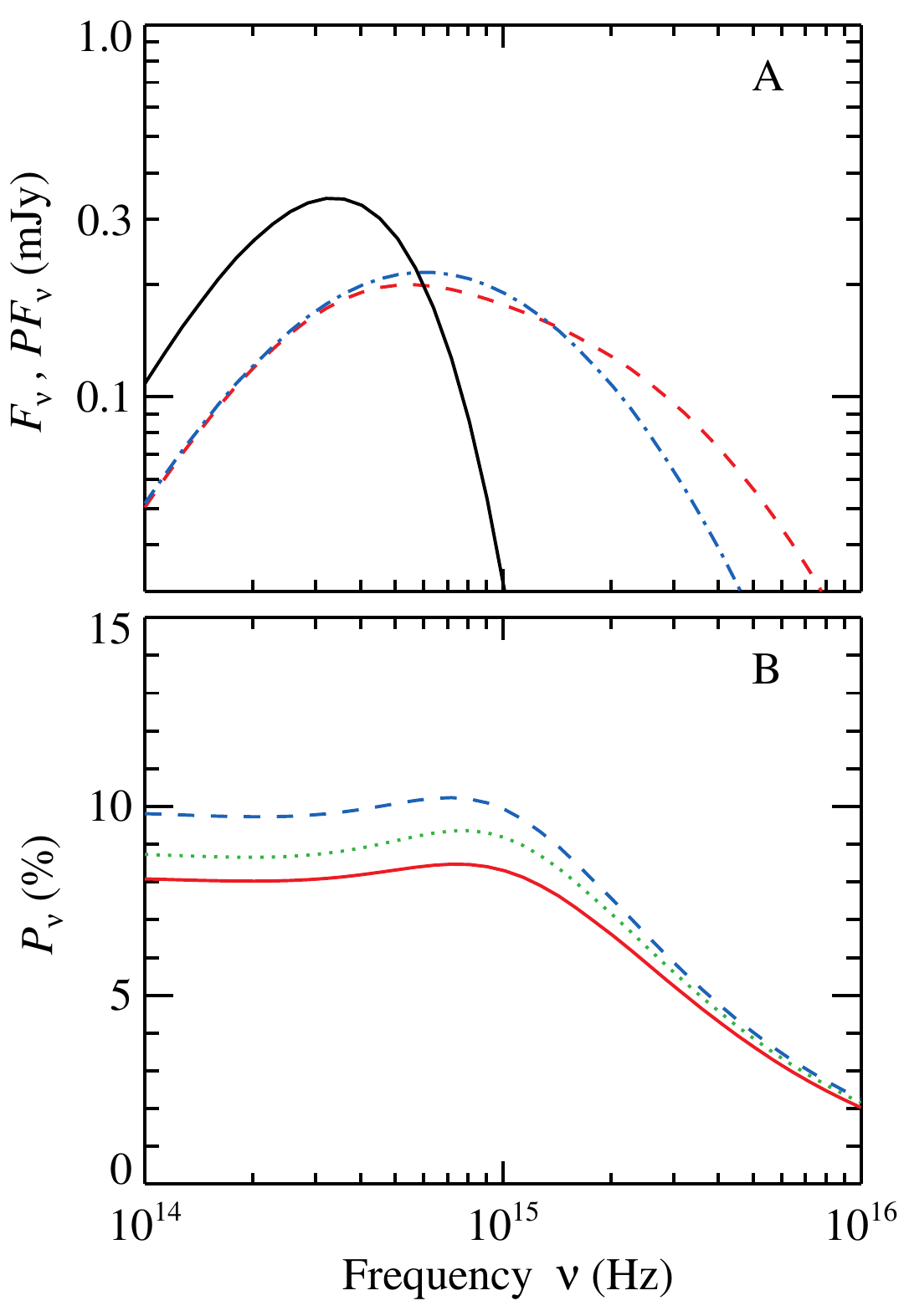}
\caption{\textbf{Polarization properties of the hot accretion flow model.}
\textbf{(A)} Flux from the spectral components: blackbody ($T_{\rm in}=6200$~K, black solid line) and first Compton scattering of these photons by hot medium (red dashed line) with $kT_{\rm e}=200$~keV and scale-height given by parameter $\cos\alpha_{\rm crit}=0.5$, and spectrum of the polarized flux of the first scattering component (blue dot-dashed line) that has been scaled up to match the low-energy part of the intensity spectrum (divided by the PD at low energies). 
This shows that the polarized flux has the same shape as the scattered component flux, justifying our joint model fitting of the UV flux together with the polarized flux (shown in Fig.~\textbf{2}).
Only results for inclination of $i_{\rm orb}=73\degr$ are shown, as the dependence on the inclination in its allowed range is minor.
\textbf{(B)} PD of the scattered component as a function of frequency is shown for different inclinations $i_{\rm orb}=81\degr$ (blue dashed line), $73\degr$ (green dotted) and $66\degr$ (red solid).  
PD for energies below the peak of the scattered component is nearly frequency-independent. 
The observed strong dependence of the PD on frequency is caused by the dilution of the polarized scattered radiation by the unpolarized accretion disk.
\label{fig:adaf}}
\end{figure}

The additional polarized component may instead arise from Compton up-scattering of soft photons by hot electrons, which is related to a systematic shift in energy.
A hot, geometrically thick accretion flow can exist in low-luminosity black hole binaries and active galactic nuclei\cite{Narayan96,YN14}.
Alternatively, a hot layer (corona) atop of the cold disk, heated by the unwinding stream of matter, could be present.
In both cases, the hot medium is expected to be optically thin, and may Compton up-scatter the photons coming from the cold accretion disk.
We calculate the spectra and polarization properties of  Compton scattering using iterative scattering method\cite{PS96}. 
We assume the spectrum of incident, non-polarized photons is a blackbody with temperature equal to the disk inner temperature $T_{\rm in}=6200$~K. 
For these seed photons, to explain the peak of the first scattering order at $\sim10^{15}$~Hz (as demanded by the spectral decomposition), we need to have electrons with temperature of about 200~keV.
For such a high temperature, the PD is expected to drop with respect to the maximal, Thomson-regime PD by a factor of 3\cite{Pou94ApJS}.
In the case of inner hot flow, only the photons travelling at inclinations $\alpha$ to the disk normal that are larger than some critical value $\alpha_{\rm crit}$ can interact with the hot matter. 
This limiting angle is related to the scale-height $H$ of the hot medium $\cot\alpha_{\rm crit} \sim  (H/R_{\rm in})$. 
Increasing the flow scale-height leads to a drop of PD.
In Fig.~\ref{fig:adaf} we show the resulting flux, polarized flux and PD spectra at different viewing angles $i_{\rm orb}=66\degr$, $73\degr$ and $81\degr$.
In this case the polarization is parallel to the disk normal.
We find that the PD of scattered radiation is nearly independent of the photon frequency up to its peak at $\nu\sim10^{15}$~Hz and therefore the spectral shape of the flux and $PF$ of the scattered component are nearly identical in the optical band.

For the slab-corona geometry, an appropriate parameter determining PD is the Thomson optical depth of the hot slab $\tau_{\rm T}$.  
Only for $\tau_{\rm T}\gtrsim 0.5$ a substantial polarization exceeding 5\% can be produced and the dominant electric-field oscillations are perpendicular to the projection of the disk normal.
However, Comptonization in a slab of  $\tau_{\rm T}=0.1$ and electron temperature $kT_{\rm e}=200$~keV overproduces the observed x-ray flux. 
This would imply that a more probable model that is more consistent with the observed spectral energy distribution and polarization properties is scattering of the disk radiation in the inner hot flow. 
Such a geometry favors the measured PA being parallel to the orbital axis. 

Finally, the blue PD spectrum could be produced by dust scattering of the accretion disk radiation. 
This process is thought to be responsible for similarly blue polarization spectra observed from supermassive black holes in Seyfert galaxies and quasars\cite{Berriman89,Brindle90,Miller91,Webb93,Goodrich94,Kishimoto08}. 
The dust would  likely be located in a flattened envelope (equatorial wedge) around the accretion disk\cite{Goodrich94},  where it is shielded from the inner disk radiation,  or in a circumbinary disk\cite{Muno06}.
In this case, the polarization vector lies in the meridional plane. 
If instead the dust has a more spherical distribution, the PD is expected to be smaller and the polarization to be perpendicular to the meridional plane.
The dust scattering model, however, does not explain the UV excess.

\subsection*{Geometry}

Here we define the coordinate systems and derive the formulae to compute the misalignment angle between the black hole spin  and the orbital angular momentum, and the azimuthal angle of the black hole spin projection in the orbital plane. 
We consider a Cartesian system with the $x-y$ plane coinciding with the orbital plane. Thus the unit vector of the orbital angular momentum is $\unitvec{\Omega}=(0,0,1)$.
We choose the direction to an observer to lie in the $x-z$ plane at inclination angle $i_{\rm orb}$ as measured from the orbital axis (see Fig.~\ref{fig:geometry_orbit} for geometry), so  the observer unit vector is
\begin{equation}
\unitvec{o} = (\sin i_{\rm orb}, 0, \cos i_{\rm orb}). 
\end{equation}
We assume that the black hole spin is directed at an angle $\beta$ from the $z$-axis at azimuthal angle $\Phi_{\rm bh}$, which is measured from the $x$-axis in the counter-clockwise direction in the $x-y$ plane as viewed from the top.
The unit vector of the black hole spin is 
\begin{equation}
\unitvec{s} = (\sin \beta \cos \Phi_{\rm bh}, \sin \beta \sin \Phi_{\rm bh}, \cos \beta). 
\end{equation}

\renewcommand{\thefigure}{{\bf S3}} 
\begin{figure}
\centering
\includegraphics[width=0.7\columnwidth]{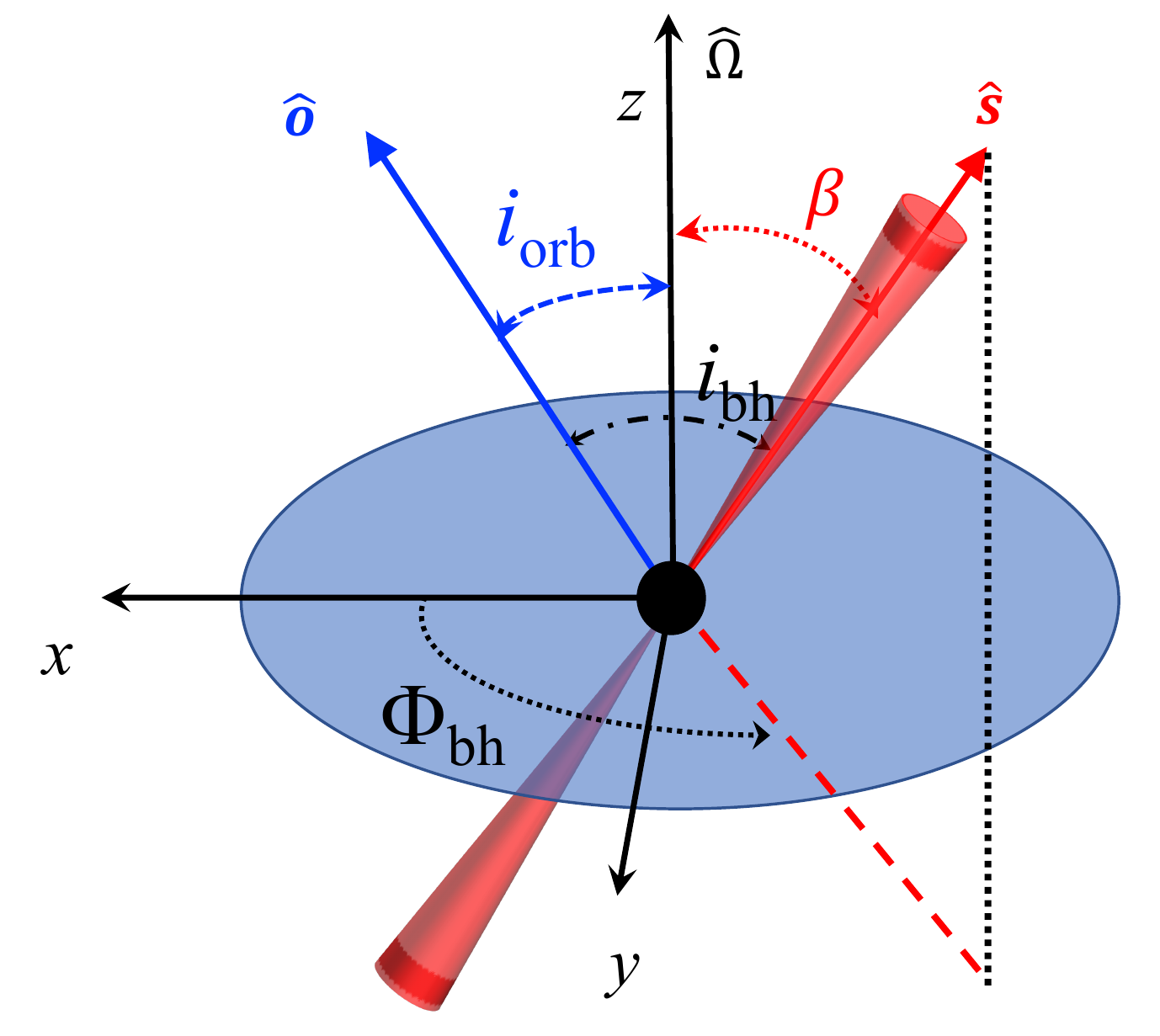}
\caption{\textbf{Geometry of the system  from the point of view of the binary.} 
The orbital plane here coincides with the $x-y$ plane of the Cartesian system with the orbital axis being along the $z$-axis. 
The observer is situated in the $x-z$ plane at inclination angle $i_{\rm orb}$ as measured from the orbital axis.  
The black hole spin is directed at angle $\beta$ from the $z$-axis at azimuthal angle $\Phi_{\rm bh}$ as measured from the $x$-axis in the counter-clockwise direction in the $x-y$ plane as viewed from the top.  
The red cones indicate the jet, and the blue disk indicates the binary orbit. 
Fig.~\textbf{3} shows the same geometry from the observer's perspective.
\label{fig:geometry_orbit}}
\end{figure}

We define the polarization basis with the unit vector $\unitvec{e}_1$ directed on the sky in the direction of the projection of the orbital spin, vector $\unitvec{e}_2$ being directed to the left on the sky, and the third vector coincides with the observer direction: 
\begin{equation}
\begin{aligned}
\unitvec{e}_1 &= \frac{\unitvec{\Omega}-\cos i_{\rm orb} \ \unitvec{o}}{\sin i_{\rm orb}} = (-\cos i_{\rm orb}, 0, \sin i_{\rm orb}), \\
\unitvec{e}_2&= (0,-1,0),  \\
\unitvec{e}_3&= \unitvec{o} .  \\
\end{aligned}
\end{equation}
In this vector basis, the black hole and the orbital spin vectors can be represented as 
\begin{eqnarray}
\unitvec{s}&=&(\sin i_{\rm bh} \cos \Delta, \sin i_{\rm bh} \sin \Delta, \cos i_{\rm bh}), \\
\unitvec{\Omega}&=&(\sin i_{\rm orb} ,0, \cos i_{\rm orb}) ,
\end{eqnarray} 
where $\Delta=\theta_{\rm bh}-\theta_{\rm orb}$ is the difference in position angles of the black hole  and the orbital spins. 
The misalignment angle $\beta$ between the black hole and the orbital axis is then given by the scalar product: 
\begin{equation}
\label{eq:cosbeta_bh}
\cos\beta = \unitvec{\Omega} \cdot \unitvec{s} = \cos i_{\rm bh} \cos i_{\rm orb} + \sin i_{\rm bh} \sin i_{\rm orb} \cos \Delta.
\end{equation}
The direction cosines of the black hole spin vector $\unitvec{s}$ in polarization basis are 
\begin{equation}
\begin{aligned}
\unitvec{s}\cdot\unitvec{e}_1&= \sin i_{\rm bh} \cos \Delta = \cos\beta \sin i_{\rm orb} -\sin\beta \cos i_{\rm orb} \cos\Phi_{\rm bh}, \\
\unitvec{s}\cdot\unitvec{e}_2&=  \sin i_{\rm bh} \sin \Delta  = - \sin\beta\sin\Phi_{\rm bh}  ,  \\
\unitvec{s}\cdot\unitvec{e}_3&= \cos i_{\rm bh} =\cos\beta\cos i_{\rm orb} + \sin \beta \sin i_{\rm orb}\cos\Phi_{\rm bh} ,  \\
\end{aligned}
\end{equation}
allowing us to obtain the azimuthal angle of the black hole spin: 
\begin{equation} \label{eq:phibh}
\begin{aligned}
\cos \Phi_{\rm bh} & =   
\frac{\sin i_{\rm orb} \cos i_{\rm bh} - \cos i_{\rm orb}\sin i_{\rm bh}\cos \Delta}{\sin \beta },  \\  
\sin \Phi_{\rm bh} & = - \frac{\sin i_{\rm bh} \sin
\Delta }{\sin \beta }.
\end{aligned}
\end{equation}  

We assume that the black hole spin is aligned with the jet. 
If the spin is directed along the southern approaching jet, then its inclination $i_{\rm bh} = i_{\rm jet}=63\degr\pm 3\degr$ and its position angle is $\theta_{\rm bh}=180\degr+ \theta_{\rm jet} =205\fdg1\pm 1\fdg4$\cite{Atri20,Bright20,Espinasse20}. 
If, on the other hand, the black hole spin points along the northern receding jet, then $i_{\rm bh} = 180\degr - i_{\rm jet}=117\degr\pm 3\degr$ and its position angle is $\theta_{\rm bh}= \theta_{\rm jet} =25\fdg1\pm 1\fdg4$.

\renewcommand{\thetable}{{\bf S5}}
\begin{table*} 
 \caption{\textbf{Geometrical parameters.}
 Geometrical parameters for the 16 possible cases identified by letters A--P of  relative orientation of the orbital and the black hole spins. 
 \label{tab:geom}}
\begin{footnotesize}
\begin{tabular}{lcp{0.cm}rp{0.cm}r p{0.cm}p{0.cm}rp{0.cm}r}     
    \hline
  & $i_{\rm bh}$ & \multicolumn{4}{c}{$i_{\rm jet}=63\degr\pm3\degr$} & & \multicolumn{4}{c}{$180\degr-i_{\rm jet}=117\degr\pm3\degr$} \\
  & $\theta_{\rm bh}$ & \multicolumn{4}{c}{$\theta_{\rm jet}+180\degr=205\fdg1\pm1\fdg4$} & & \multicolumn{4}{c}{$\theta_{\rm jet}=25\fdg1\pm1\fdg4$} \\
  \cline{3-6} 
   \cline{8-11} 
  & $i_{\rm orb}$ &  & $73\degr\pm6\degr$  & &  $107\degr\pm6\degr$  & & &  $73\degr\pm6\degr$  & &  $107\degr\pm6\degr$ \\
\hline
 \multicolumn{11}{c}{Polarization parallel to the meridional plane}     \\ 
 $\theta_{\rm orb}=\langle \mbox{PA}\rangle$ 
& $\beta$ (deg)   & \multirow{2}{*}{A\!\!\!\!} & $117.3\pm4.3$ & \multirow{2}{*}{B} & $137.6\pm2.4$ & &  \multirow{2}{*}{C} & $ 62.7\pm4.3$ & \multirow{2}{*}{D} &  $42.4\pm2.4$ \\
=$-19\fdg7\pm1\fdg2$ 
   & $\Phi_{\rm bh}$ (deg) & & $45.2\pm3.4$ & & $ 69.7\pm7.0$ &  & & $225.2\pm3.4$ & & $249.0\pm7.0$ \\
& & & &   &  & & &   &  & \\ 
\multirow{2}{*}{$\theta_{\rm orb}=\langle \mbox{PA}\rangle$+180\degr}& $\beta$  (deg)  &\multirow{2}{*}{E} & $ 42.4\pm2.4$ &\multirow{2}{*}{F} & $ 62.7\pm4.3$ & & \multirow{2}{*}{G} & $137.6\pm2.4$ & \multirow{2}{*}{H} &$117.3\pm4.3$ \\
   & $\Phi_{\rm bh}$ (deg)  & & $289.8\pm7.0$ & & $314.8\pm3.4$ & & & $110.3\pm7.0$ & & $134.8\pm3.4$ \\
& & & &  & & & & &   & \\ 
\hline
 \multicolumn{11}{c}{Polarization perpendicular to the meridional plane}     \\ 
\multirow{2}{*}{$\theta_{\rm orb}=\langle \mbox{PA}\rangle$+90\degr}  & $\beta$ (deg) &\multirow{2}{*}{I} & $117.0\pm4.3$ & \multirow{2}{*}{J} & $137.2\pm2.4$ & & \multirow{2}{*}{K} & $63.0\pm4.3$ &\multirow{2}{*}{L} & $42.8\pm2.4$ \\
   & $\Phi_{\rm bh}$ (deg) & & $314.5\pm3.4$ & & $290.3\pm6.9$ & & & $134.5\pm3.4$ & & $110.3\pm6.9$ \\
& & & & &  & & & & &  \\ 
\multirow{2}{*}{$\theta_{\rm orb}=\langle \mbox{PA}\rangle$+270\degr} & $\beta$  (deg)  &\multirow{2}{*}{M} & $42.8\pm2.4$ & \multirow{2}{*}{N} & $ 63.0\pm4.3$ && \multirow{2}{*}{O} & $137.3\pm2.4$ &\multirow{2}{*}{P} & $117.0\pm4.3$ \\
   & $\Phi_{\rm bh}$ (deg) & & $69.7\pm6.9$ & &  $45.5\pm3.4$ & & & $249.7\pm6.9$ &&  $225.5\pm3.4$ \\ 
       \hline
\end{tabular}
\end{footnotesize}
\end{table*}

The polarimetric data provide us with the average  polarization angle $\langle \mbox{PA}\rangle=-19\fdg7\pm1\fdg2$, which carries information about orientation of the orbital axis on the sky. 
If dominant oscillations of the electric field lie in the meridional plane formed by the orbital spin and photon propagation direction, the position angle of the orbital spin can be either $\theta_{\rm orb}=\langle \mbox{PA}\rangle$ or $\langle \mbox{PA}\rangle+180\degr$.
The electric field oscillations can also be  perpendicular to the meridional plane, then the orbital spin position angle is $\langle \mbox{PA}\rangle+90\degr$ or $\langle \mbox{PA}\rangle+270\degr$.  
Furthermore, the radial velocity measurements are not able to differentiate between inclinations $i_{\rm orb}$ and $180\degr-i_{\rm orb}$. 
The possible combinations result in 16 different geometrical arrangements of the black hole and orbital spins that satisfy the observational constraints. 
These 16 cases can be reduced to eight different values  for the misalignment angle $\beta$:   four for misalignment less than $90\degr$ and four for misalignment between $90\degr$ and $180\degr$ for the retrograde rotation of the black hole (Table~\ref{tab:geom}).

The probability distribution for the orbital inclination $i_{\rm orb}$ was assumed to be a Gaussian with the peak at 73\degr\, with $1\sigma$ error of 6\degr\,  and a cutoff at 81\degr\cite{Torres19,Torres20}. 
For an alternative case of inclination exceeding 90\degr, the distribution mirror reflected  relative to  90\degr\, is considered. 
Other parameters are assumed to follow a Gaussian distribution with corresponding $1\sigma$ errors. Using Monte-Carlo simulations\cite{Poutanen22}, we obtain the probability distributions for $\beta$ and $\Phi_{\rm bh}$ using Equations (\ref{eq:cosbeta_bh}) and (\ref{eq:phibh}), respectively. 
Their mean and standard deviation are given in Table~\ref{tab:geom}. 
In Fig.~\ref{fig:distr_beta_all} we show the posterior probability distribution for $\beta$ for the eight different cases from Table~\ref{tab:geom}. 
The probability distributions for $\Phi_{\rm bh}$  for the 16 cases from Table~\ref{tab:geom} are shown in Fig.~\ref{fig:jet_azimuth}. 

\renewcommand{\thefigure}{{\bf S4}} 
\begin{figure}
\centering
\includegraphics[width=0.9\columnwidth]{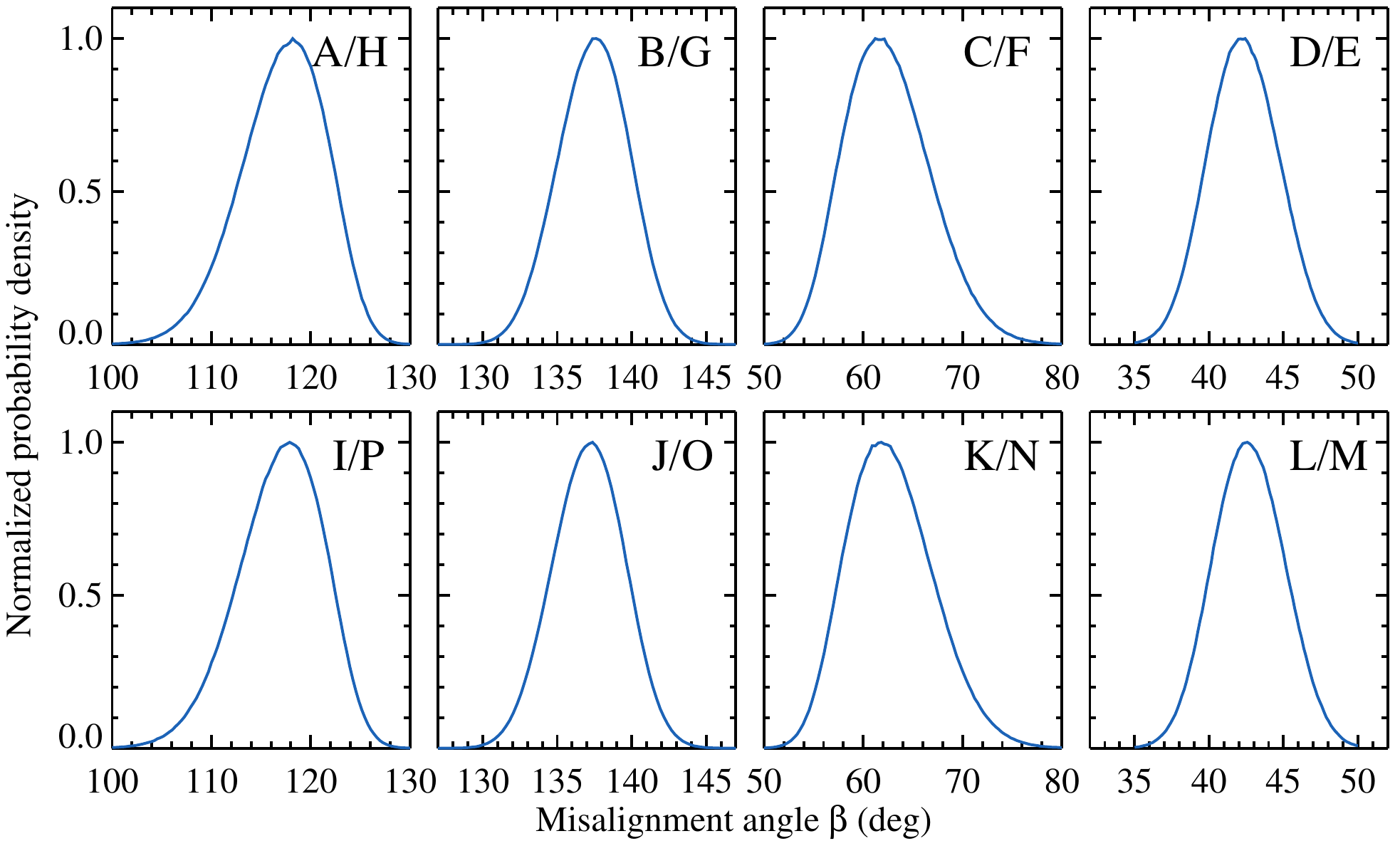}
\caption{\textbf{Probability distribution function for the misalignment angle.} 
Distributions normalized to the peak values are shown for the eight different cases presented in Table~\ref{tab:geom}. 
\label{fig:distr_beta_all}}
\end{figure}

\renewcommand{\thefigure}{{\bf S5}} 
\begin{figure}
\centering
\includegraphics[width=0.9\columnwidth]{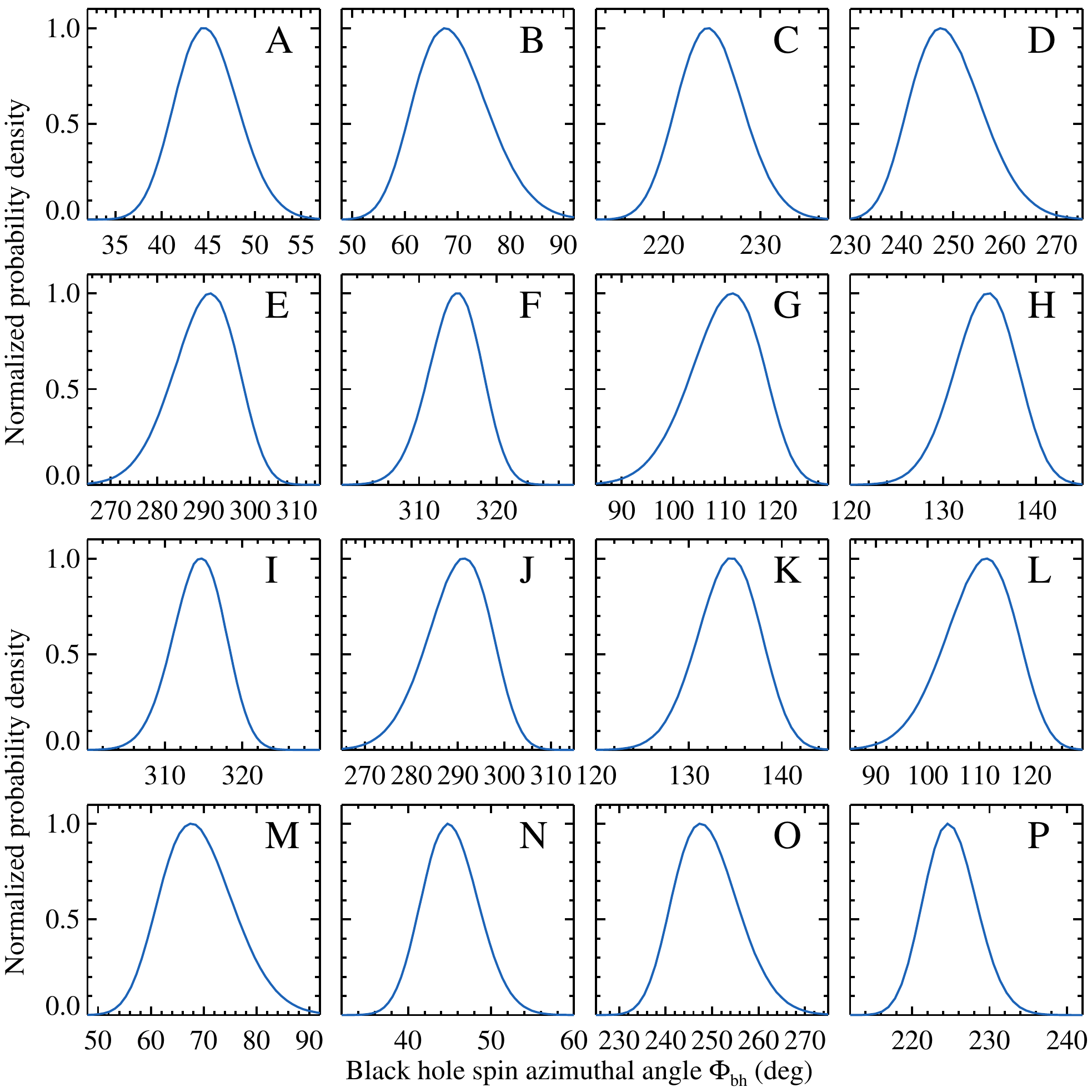}
\caption{\textbf{Probability distribution function for the black hole spin azimuthal angle.}
Distributions normalized to the peak values are shown for the 16 different cases \textbf{(A-P)} presented in Table~\ref{tab:geom}.  
\label{fig:jet_azimuth}}
\end{figure}

 \clearpage

\renewcommand\refname{\large References and Notes}



\end{document}